\documentclass[10pt,english,english,nofootinbib,notitlepage,superscriptaddress,aps,prd]{revtex4-1}

\usepackage{enumerate}
\usepackage{amsmath}
\usepackage{amsfonts}
\usepackage{amssymb}
\usepackage[utf8]{inputenc}
\usepackage[T1]{fontenc}
\usepackage{mathtools}
\usepackage{wasysym}
\usepackage{accents}
\usepackage[svgnames]{xcolor}
\usepackage[colorlinks=true, pdfstartview=FitV, linkcolor=blue, citecolor=red, urlcolor=magenta]{hyperref}
\usepackage{url}
\usepackage{graphicx}
\usepackage{subfig}
\usepackage{bm}
\usepackage{tikz}
\usetikzlibrary{arrows,decorations.markings}
\usepackage{csquotes}
\newtheorem{Definition}{Definition}[section]
\usepackage{changepage}

\begin{document}
\newcommand{\Areia}{
\affiliation{Department of Chemistry and Physics, Federal University of Para\'iba, Rodovia BR 079 - Km 12, 58397-000 Areia-PB,  Brazil.}
}
\newcommand{\Lavras}{
\affiliation{Physics Department, Federal University of Lavras, Caixa Postal 3037, 37200-000 Lavras-MG, Brazil.}
}

\newcommand{\JP}{
\affiliation{Physics Department, Federal University of Para\'iba, Caixa Postal 5008, 58059-900, Jo\~ao Pessoa, PB, Brazil.}
}

\title{Quantum Configuration and Phase Spaces: Finsler and Hamilton Geometries}

\author{Saulo Albuquerque}\email{saulosoaresfisica@gmail.com}
\JP

\author{Valdir B. Bezerra}\email{valdir@fisica.ufpb.br}
\JP

\author{Iarley P. Lobo}\email{lobofisica@gmail.com}
\Areia
\Lavras

\author{Gabriel Macedo}\email{gabriel_macsa2012@hotmail.com}
\JP

\author{Pedro H. Morais}\email{phm@academico.ufpb.br}
\JP

\author{Ernesto Rodrigues}\email{dota.jp13@gmail.com}
\JP

\author{Luis C. N. Santos}
\email{luis.santos@ufsc.br}
\JP

\author{Gislaine Varão}\email{gislainevarao@gmail.com}
\JP

\date{\today}

\begin{abstract}
In this paper, we review two approaches that can describe, in a geometrical way, the kinematics of particles that are affected by Planck-scale departures, named Finsler and Hamilton geometries. By relying on maps that connect the spaces of velocities and momenta, we discuss the properties of configuration and phase spaces induced by these two distinct geometries. In particular, we exemplify this approach by considering the so-called $q$-de Sitter-inspired modified dispersion relation as a laboratory for this study. We finalize with some points that we consider as positive and negative ones of each approach for the description of quantum configuration and phases spaces.
\end{abstract}

\pacs{}
\maketitle


\section{Introduction}

Since the original works by Bronstein~\cite{Bronstein:2012zz} that demonstrated  uncertainty in the localization of events when geometrical degrees of freedom are quantized, it has been argued that attempts to formulate quantum gravity in a differentiable manifold endowed with smooth geometric quantities would not be an interesting path to follow if one aims to pursue a fundamental approach to this problem. Attempts in this direction have accumulated over the years, having prominent representatives such as loop quantum gravity (LQG)~\cite{Rovelli:2008zza} and causal dynamical triangulation~\cite{Ambjorn:2010rx}. These approaches to quantum gravity predict or describe several effects that should be manifest at the Planckian regime of length and energy, such as the discretization of geometry, which requires a language that obviously departs from the usual Riemannian construction of general relativity. Despite the elegance of such approaches, with current technology we are far from being able to concretely address the regime in which such discretization would become evident. Nevertheless, the notion that spacetime could effectively behave like a medium formed by 
''atoms of space'' 
has led to a rich phenomenological approach to quantum gravity, which by encoding generic departures from relativistic equations, can describe common predictions expected to be present at an intermediate stage between classical and quantum gravity. Such an approach is encompassed in the {area} of quantum gravity phenomenology, which addresses a myriad of effects beyond the one described in this paragraph, as can be seen 
in~Ref. 
\cite{Amelino-Camelia:2008aez}, and in particular, has found in multimessenger astronomy a fruitful environment to be explored~\cite{Addazi:2021xuf}.
\par
Usually, the regime,
in which this idea is considered,  
is the regime, in which the test particle approximation is valid consisting 
of the approximation, in which one would 
have simultaneously faint gravitational and quantum effects, described by the limits
 of the gravitational constant,
$G\rightarrow 0$, and the reduced Planck's constant,
$\hbar\rightarrow 0$, however, with the Planck energy, $E_P=\sqrt{c^5\hbar/G}$, being 
finite, with $c$ the speed of light. 
This deformed 
''Minkowski limit,'' 
which presents departures from Minskowski spacetime's structure has been suggested by various quantum gravity proposals, such as the linearization of the hypersurface deformation algebra inspired by LQG~\cite{Amelino-Camelia:2016gfx,Brahma:2016tsq,Brahma:2018rrg} and non-commutative geometry~\cite{Majid:1994cy,Lukierski:1991pn,Lukierski:1992dt,Majid:1996kd} (for more details on this Miskowski limit, 
see 
Section 3.1.1 of~Ref. 
\cite{Amelino-Camelia:2008aez}, and for more references on other theoretical approaches in which such limit emerges, please refer to Section 2.2 
of~Ref. 
\cite{Addazi:2021xuf}). It is expected that the path between the differentiable Riemannian description of special (and general) relativity and the complete quantum gravity theory should pass through an intermediate 
regime, 
in which  
one has  
departures from the Riemannian character of spacetime but still 
 has  
geometric features that could describe a bottom-up phenomenology.
\par
Furthermore, geometry plays an important role in the description of principles that have guided the developments of relativistic theories; for example, the principle of covariance is manifest through the use of tensorial equations of motion, the local relativity principle is a physical manifestation of having local equations of motion invariant under the Poincar\'e group (which is the group of isometries of Minkowski space), the equivalence principle of general relativity is manifest in the fact that the motion of free particles is realized through geodesics, and the clock postulate can be expressed by stating that an observer measures its proper time by the 
arc-length of its own trajectory.
\par
An important part of quantum gravity phenomenology is devoted to the question of whether, in the aforementioned Minkowski limit, the Lorentz invariance, and consequently, the local relativity principle, is preserved or broken due to Planck-scale effects~\cite{Amelino-Camelia:2010lsq}. As is known, a length/energy scale is not invariant under Lorentz transformations, which implies that either a quantum gravity scale breaks the equivalence of inertial frames in the aforementioned Minkowski limit, or the Lorentz or Poincar\'e group only describes a low energy/large distance approximation of a deeper transformation between inertial frames. The former possibility is known as a Lorentz invariance violation (LIV) scenario~\cite{Mattingly:2005re,Liberati:2013xla}, and the latter is known as doubly (or deformed) special relativity (DSR)~\cite{Amelino-Camelia:2000stu,Magueijo:2001cr}. As the geometrization of special relativity, due to Minkowski, paved the way to more fundamental descriptions of nature, we shall follow a similar path, but of geometrizing DSR.
\par
Geometric descriptions of DSR through non-commutative geometry are known \cite{Majid:1994cy,Lukierski:1991pn,Lukierski:1992dt,Majid:1996kd}, but we revise some continuous, differentiable ways of exploring non-Riemannian degrees of freedom and the possibilities for preserving the aforementioned principles. This way, we critically analyze two extensions of Riemannian geometry that are capable of describing aspects of an emergent 
''quantum configuration and phase spaces''  
that preserve the intuition of those principles: they are Finsler and Hamilton geometries. Finsler 
geometry originally {is related} to the space of events and velocities (for this reason we refer to a quantum configuration space), and Hamilton geometry originally described the space of events and momenta (for this reason, we call it a quantum phase space). In this paper, we revise the phenomenological opportunities that emerge from these approaches and the interplay between them. We also condensate the utility of each of these geometries and their limitations in the current scenario.
\par
{We should also stress that the approaches
described  
 in this review, refer to configuration and phase spaces probed by a single particle. The geometry probed by a multi-particle system and its interplay with Finsler and Hamilton languages (or even geometries that go beyond them) should still be further explored, in which, possibly the intuition gained from the 
relative locality
framework~\cite{Amelino-Camelia:2011lvm} would play a prominent role in this approach.}
\par
 
The
paper is organized as follows. 
Section~\ref{sec:rainbow} revisits
 the origin of the idea of describing the effective spacetime probed by a particle that propagates through a modified dispersion relation (MDR) by the proposal of rainbow metrics. 
 
Section~\ref{sec:finsler} 
 revisits
how this general idea is realized by the use of Finsler geometry in the tangent bundle, whose dual version in the cotangent bundle is discussed in Section~\ref{sec:dualfinsler}, which 
 is illustarated  
by considering the curved non-trivial case of $q$-de Sitter-inspired Finsler geometry. 
Section~\ref{sec:hamilton} 
 considers 
the situation of deriving the geometry of the cotangent bundle, and,   
in Section~\ref{sec:dualhamilton}, 
its dual tangent bundle formalization defined by Hamilton geometry is considered, 
which is illustrated by   
 the $q$-de Sitter case.  
In Section~\ref{sec:advdiff}, we comparatively discuss these two approaches and highlight points that we 
consider as useful
 as well as 
their limitations. Finally,  some important 
remarks are drawn 
in Section~\ref{sec:conc}. 
 Throughout the paper, 
a 
system of units with 
$c=\hbar=1$ is used, 
  so 
that the Planck length is the inverse of the Planck energy: 
$\sqrt{G}=\ell=E_{\text{P}}^{-1}$. 

\section{Preliminaries on Rainbow Geometries}\label{sec:rainbow}
As described above, 
over the years, the intuition that spacetime would behave like material media, where instead of atoms of matter, one would have atoms of spacetime, has been solidified through some approaches of quantum gravity. Just as occurs in matter, in which one does not need to know the specific details of the granular structure of a given medium to study the propagation of particles through it, in spacetime, 
one  
can build phenomenology-inspired ways of modeling how elementary particles interact with discrete gravitational degrees of freedom while traveling through space, a so-called 
''in-vaccuum dispersion.'' 
One  
could say that the most popular way of doing this is through the assumption that particles would obey a modified dispersion relation, whose 
corrections are given perturbatively by powers of the quantum gravity scale, which we could assume as being in the order of Planck units. The dispersion relation furnishes the group velocity of waves and defines the trajectory that on-shell particles follow from the Hamilton equations. 
Actually, when the interplay between the presence of amplifiers of observables and the uncertainties of observations allows us to constrain this parameter at a level close to its Planckian version, we say that we are at Planck-scale sensitivity~\cite{Amelino-Camelia:2008aez}.
\par
Such behavior also happens in meta-materials~\cite{proutorov2018finsler}, in which it is possible to describe the motion of particles through it 
by geodesics in a given geometry; it also appears in the motion of a charged particle in a pre-metric formulation of 
electromagnetism~\cite{hehl2003foundations}, in the description of seismic waves~\cite{vcerveny2002fermat}, etc.; for a review,  
see~Ref. \cite{Pfeifer:2019wus}. 
Additionally, one could wonder if the motion of particles,  
determined by Planck-scale modified dispersion relations,  
could also be described by geodesics of a non-Riemannian geometry. Besides, the dispersion relation itself is usually determined by the norm of the $4$-momentum measured by a Riemannian metric, which also determines the symmetries observed by measurements in that spacetime.
\par
This intuition was early realized by the so-called 
''rainbow geometries''~\cite{Magueijo:2002xx}, idealized by Jo\~ao Magueijo and Lee 
Smolin  
 which aimed to extend the DSR formulation proposed by them 
in~Ref.
\cite{Magueijo:2001cr} to curved spacetimes. In that case, the way found 
to express local modified dispersion relations through a norm,  
consisted in absorbing functions of the particle's energy divided by 
 Planck 
energy, {$\epsilon=E/E_{\text{P}}$}, such as $f(\epsilon)$ and $g(\epsilon)$, which would appear in the MDR that follows:
\begin{equation}
    m^2=f^2(\epsilon)E^2-g^2(\epsilon)|\vec{p}|^2\, ,
\end{equation}
(with the three-momentum $\vec{p}$) into the definition of new spacetime tetrads, $\tilde{e}_{(0)}^{\quad 
\mu}(\epsilon)=f(\epsilon)e_{(0)}^{\quad\mu}$ and $\tilde{e}_{(I)}^{\quad\mu}(\epsilon)=g(\epsilon)e_{(I)}^{\quad \mu}$, such that the MDR reads
\begin{equation}
    m^2=\eta^{AB}\tilde{e}_{(A)}^{\quad\mu}\tilde{e}_{(B)}^{\quad\nu}p_{\mu}p_{\nu}=\tilde{g}^{\mu\nu}(\epsilon)p_{\mu}p_{\nu}\, ,
\end{equation}
where $g^{\mu\nu}(\epsilon)=\eta^{AB}\tilde{e}_{(A)}^{\quad\mu}\tilde{e}_{(B)}^{\quad\nu}$ is the rainbow metric, 
$\eta^{AB}$ is {the} 
Minkowski metric $\text{diag}(+---)$ , Greek letters denote four-dimensional indices and take on the values 0 (time) 1, 2, and 3 (space), 
 low-case Latin letters denote the space indices, and $p_{\mu}$ is the 4-momentum.  
This description would imply that when an observer uses the motion of a particle with energy $E$ to probe spacetime, then the line element assigned to that spacetime is the following:
\begin{equation}\label{rainbow-line-element}
    ds^2=\tilde{g}_{\mu\nu}dx^{\mu}dx^{\nu}=\frac{g_{00}}{f^2(\epsilon)}(dx^{0})^2+\frac{g_{ij}}{g^2(\epsilon)}dx^idx^j+2\frac{g_{0i}}{f(\epsilon)g(\epsilon)}dx^0dx^i\, ,
\end{equation}
where $g_{\mu\nu}$ is the metric found from undeformed tetrads. Thus, in a nutshell, one identifies the rainbow functions, $f$ and $g$,
from a MDR that is usually inspired by fundamental theories of quantum gravity or by phenomenological intuition; then, one uses $\tilde{g}_{\mu\nu}$ as an input into the classical gravitational field equations. Considering modifications of the stress-energy tensor due to the rainbow functions, one derives what should be $g_{\mu\nu}$ (since $f$ and $g$ are determined a priori). Usually, this procedure gives that $g_{\mu\nu}$ is the Riemannian metric found from the usual gravitational field equations. Therefore, this approach gives basically the usual metric components of a given theory, just modified by factors of the rainbow functions as 
in Equation 
 \eqref{rainbow-line-element}.
\par
Effective energy-dependent spacetimes have emerged in different approaches to the description of the quantization of gravitational/geometric degrees of freedom~\cite{Weinfurtner:2008if,Assanioussi:2014xmz,Olmo:2011sw}. Along this line of research, Magueijo-Smolin's proposal has been applied in a myriad of contexts, such as in black hole physics~\cite{Ling:2005bp,Lobo:2021bag}, cosmology~\cite{Gorji:2016laj}, wormholes~\cite{Garattini:2015pmo,Amirabi:2018ncf}, cosmic strings~\cite{Bezerra:2019vrz}, disformal geometries~\cite{Carvalho:2015omv,Lobo:2017bfh}, and electrostatic self-interaction of charged particles~\cite{Santos:2019}. However, despite its range of applicability and utility in furnishing intuition about extreme scenarios, this approach presents some conceptual and technical limitations that seem unavoidable, such as the lack of a rigorous mathematical framework in which this idea is formulated or the imposition of a preferred vielbein in which the particle's energy is measured, which seems in contradiction with the local DSR intention of this proposal. As 
 shown below, 
the solution to these problems is actually coincident, and the search for a rigorous mathematical formulation for these ideas will be responsible for giving a 
framework,  
in which proper physical questions 
can be answered 
and novel phenomenological opportunities to 
 born. The main issue here is what is the proper formulation of a geometry that should not only depend on spacetime points, but also should carry 
energy dependence of the particle itself that probes this spacetime. 
This
paper 
 deals  
with the 
two main proposals---Finsler and Hamilton geometries---
solving  
 some of the raised 
 problems  
and
also discusses 
limitations on their owns. 
\section{Geometry of the Tangent Bundle: Finsler Geometry}\label{sec:finsler}

The 1854 Habilitation Dissertation by Bernhard Riemann presents the germ of the idea behind what would later be called Finsler geometry. In the 
second part of the dissertation, it is said (see 
Ref. 
\cite{riemann}, p. 35): 

\begin{displayquote}
''For Space, when the position of points is expressed by rectilinear co-ordinates, $ds = \sqrt{\sum (dx)^2}$; Space is therefore, included in this simplest case. The next case in simplicity includes those manifoldnesses in which the line-element may be expressed as the fourth root of a quartic differential expression. The investigation of this more general kind would require no really different principles, but would take considerable time and throw little new light on the theory of space, especially as the results cannot be geometrically expressed; I restrict myself, therefore, to those manifoldnesses in which the line-element is expressed as the square root of a quadric differential 
expression.''
\end{displayquote}

The exploration of such more general cases of line elements 
 will 
be done only 64 years later,  in 1918, 
in the Ph.D. thesis of Paul Finsler~\cite{book:1130902},
 where  
at least from the metric point of view, the distance between points is 
(''distance''), please confirm.
measured by a 1-homogeneous function 
 (homogeneous with the degree of 1) 
Such a metric tensor would be defined in the tangent bundle of the base manifold, since it would depend not only on the manifold points, but also on a direction, which is a manifestation of the non-Pythagorean nature of this space. Later on, the issue of non-linear connections was further developed and incorporated as a fundamental structure for the dynamical description of Finsler spaces (for a historical perspective on Finsler geometry, we refer the reader to the Preface 
of~Ref. 
\cite{bao2000introduction} and references therein). The case of pseudo-Finsler geometries, as an arena for describing spacetime, has been recently discussed~\cite{Hohmann:2021zbt,Bernal:2020bul}, 
where,
for instance, different definitions are presented and important theorems regarding its causal structure among other issues are being derived~\cite{Minguzzi:2014aua}.
\par
In Section \ref{sec:rainbow}, 
a glimpse of the non-Riemannian nature of spacetime was notified 
emerging 
 as a manifestation of the quantization of gravitational degrees of freedom. 
 Actually, 
 as one
can anticipate, the non-quadratic, i.e., 
non-Pythagorean nature of a dispersion relation is connected to a possible Finslerian nature of spacetime through an intermediate step that connects the kinematics of particles in a Hamiltonian to a Lagrangian formulation. 
 Actually, 
the MDR corresponds 
to a Hamiltonian constraint, which 
physical particles supposedly obey, 
the way  
 that the trajectories of free 
particles, induced by such a deformed Hamiltonian, capture the propagation of a particle through a quantized spacetime. 
For this reason, the Helmholtz action, associated with such a particle, is naturally given by the functional,
\begin{equation}\label{action-ham1}
    S[x,p,\lambda]=\int d\mu (\dot x^{\alpha} p_{\alpha} - \lambda f(H(x,p),m))\, ,
\end{equation}
where the dot denotes 
 differentiation with respect to the parameter $\mu$, $p_{\mu}$ is the particle's momenta, $f$ is a function that is null if the dispersion 
relation is satisfied, namely, $H(x,p)=m$, and $\lambda$ is a Lagrange multiplier. This is a premetric formulation that is actually defined in the space $T^*M\times {\mathbb R}$, where $T^*M$ is the phase space of analytical mechanics or cotangent bundle. In order to find 
an arc-length, and consequently, a geometric structure, 
 one needs  
to calculate an equivalent Lagrangian defined in the {configuration space} or the tangent bundle $TM$ described by points and velocities (such an observation was firstly presented 
in~Ref. 
\cite{Girelli:2006fw}). The algorithm for doing so is 
 as follows~\cite{Lobo:2020qoa}:
\begin{enumerate}
    \item variation 
with respect to $\lambda$ enforces the dispersion relation; 
    \item variation 
with respect to $p_{\mu}$ yields an equation $\dot x^{\mu} = \dot x^{\mu}(p,\lambda)$, 
which must be inverted to obtain $p_{\mu}(x,\dot x, \lambda)$ to eliminate the momenta  $p_{\mu}$ from the 
action; 
    \item using 
$p_{\mu}(x,\dot x, \lambda)$ in the dispersion relation, one can solve for $\lambda(x,\dot x)$; and 
    \item {finally,} 
the desired length measure is obtained as $S[x] =  S[x,p(x,\dot x,\lambda(x,\dot x)),\lambda(x,\dot x)]_{H}$.
\end{enumerate}

This is a Legendre transformation, whose conditions of existence and capability of providing a physical framework are discussed in 
Refs.~\cite{Raetzel:2010je,Rodrigues:2022mfj}. These formal conditions are always guaranteed when one considers deformations at the 
perturbative 
level. This is crucial because the following algorithm cannot be applied in practice if it is not possible to invert the velocity function to find the momenta as a function of the other variables. In general, this cannot be done, especially for complicated dispersion relations, such as those that depend on sums of hyperbolic functions~\cite{Gubitosi:2011hgc}. Anyway, since quantum gravity phenomenology is usually concerned with first order effects, which are those attainable by experiments nowadays, we shall concentrate on the perturbative level in order to derive our conclusions.

For example, if this algorithm is applied to a Hamiltonian of the form, 
\begin{equation}
     H(x,p) = g(p,p) + \varepsilon h(x,p)\, ,
\end{equation}
where $g(p,p)= g^{ab}(x)p_a p_b$ is an undeformed dispersion relation, $h(x,p)$ is a function of spacetime points and momenta that depends on the model under consideration, and $\varepsilon$ is the perturbation parameter that is usually a function of the energy scale of the deformation (such as the Planck or quantum gravity length scale). As shown 
in~Ref. 
\cite{Lobo:2020qoa}, after the Legendre transformation, the equivalent action takes the form, 
\begin{equation}
    S[x] = m\int d\mu \sqrt{g(\dot x, \dot x)}\left(1 -\varepsilon \frac{h(x,\bar p(x,\dot x))}{2m^2} \right)\, ,
\end{equation}
where 
$\bar p_{a}(x,\dot x) = m 
{\dot x_a}/{\sqrt{g(\dot x, \dot x)}}$. 
In particular, when $h$ is a polynomial function of momenta as (
the index is shifted: $n\rightarrow n+2$, in comparison 
 with~{Ref.} \cite{Lobo:2020qoa}, 
such that now $n$ corresponds to the power of Planck length in the MDR), 
\begin{equation}
    h(x,p) = h^{\mu_1 \mu_2 ....\mu_{n+2}}(x)p_{\mu_1} p_{\mu_2}...p_{\mu_{n+2}}\, ,
\end{equation}
and $\varepsilon=\ell^{n}$, 
 one finds 
an action of the form, 
\begin{equation}
  S[x] =   m\int d\mu \sqrt{g(\dot x, \dot x)}\left(1 - (\ell m)^{n} \frac{h_{\mu_1 \mu_2 ....\mu_{n+2}}(x)\dot x^{\mu_1} \dot x^{\mu_2} ...\dot x^{\mu_{n+2}}}{2g( \dot x,\dot x)^{\frac{n+2}{2}}}\right)\, ,
\end{equation}
where we lowered the indices of $h$ with the components of $g$. The connection between the mechanics of free particle and geometry takes place when 
the above expression is identified 
with the arc-length functional, 
$s[x]$,  
of a given geometry, i.e., $s[x]=S[x]/m$. Such an identification makes sense if we want to state that the trajectories of free particles are extremizing curves or geodesics in a given geometry, it is related to the preservation of the equivalence principle even in this Planck-scale deformed scenario.
\par
In this case, the spacetime in which a particle propagates by a MDR is described by an arc-length functional that generalizes the one of Riemannian geometry and is given by a function $F(x,\dot x)$ that is 
1-homogeneous in the velocity $\dot x$, such that the 
arc-length  
is indeed parametrization invariant, as it must be:
\begin{equation}\label{arclength1}
    s[x]=\int F(x,\dot x)d\mu\, .
\end{equation}

Actually,  
this is the kind of scenario envisaged by Riemann in his dissertation, and explored by Finsler, that emerges here quite naturally. There are some definitions of a pseudo-Finsler spacetime in the literature, but we rely on 
that given 
in~Ref.
\cite{Hohmann:2021zbt} (the differences in comparison to other definitions are discussed in Ref.  \cite{Hohmann:2021zbt}). 
First of all, we are going to work with a smooth manifold, $M$, 
endowed with a real valued positive function $L$ that takes values on the tangent bundle $TM$, described by coordinates $(x,y)$, where $\{x^{\mu}\}$ are spacetime coordinates and $\{y^{\mu}\}$ refer to vector or velocity coordinates. Actually, we shall need the slit tangent bundle $\widetilde{TM}=TM/\{0\}$, in which we remove the zero section, and we also need the projection $\pi:TM\rightarrow M$. A conic subbundle is a submanifold ${\cal D}\subset \widetilde{TM}$ such that $\pi({\cal D})=M$ and with the conic property that states that if $(x,y)\in {\cal D}\Rightarrow (x,\lambda y)\in {\cal D}, \, \forall \lambda >0$.
\par
In a nutshell, a Finsler spacetime is a triple $(M,{\cal D}, L)$, where $L:{\cal D}\rightarrow {\mathbb R}$ is a smooth function satisfying the 
conditions:  
\begin{enumerate}
    \item 
     positive 
2-homogeneity: $L(x,\alpha y)=\alpha^2 L(x,y),\,  \forall \alpha>0$; 
    \item at 
any $(x,y)\in {\cal D}$ and in any chart of $\widetilde{TM}$, the following Hessian (metric) is non-degenerate:
    \begin{equation}\label{f-metric1}
        g_{\mu\nu}(x,y)=\frac{1}{2}\frac{\partial^2 }{\partial y^{\mu}\partial y^{\nu}}L(x,y)\, ; 
    \end{equation}
    \item the 
metric $g_{\mu\nu}$ has a Lorentzian signature.
\end{enumerate}

The function $L$ is actually the square of the Finsler function, $L(x,y)=F^2(x,y)$, and from it 
the Finsler 
arc-length is defined 
as given in 
Equation 
\eqref{arclength1}. 
Condition $1$ above guarantees that Equation 
\eqref{arclength1} does not depend on the parametrization used to describe the curve and that using Euler's theorem for homogeneous functions, this expression can be cast as
\begin{equation}
    s[x]=\int \sqrt{g_{\mu\nu}(x,\dot x)\dot{x}^{\mu}\dot{x}^{\nu}}d\mu\, .
\end{equation}

From a coordinate transformation, 
\begin{align}
    \tilde{x}^{\mu}=\tilde{x}^{\mu}(x)\, ,\label{coordtrans1}\\
    \tilde{y}^{\mu}=\frac{\partial \tilde{x}^{\mu}}{\partial x^{\nu}}y^{\nu}\, ,\label{vectrans1}
\end{align}
the functions $g_{\mu\nu}$ transform according to
\begin{equation}\label{metric-transf}
    \tilde{g}_{\mu\nu}(\tilde{x},\tilde{y})=\frac{\partial x^{\alpha}}{\partial \tilde{x}^{\mu}}\frac{\partial x^{\beta}}{\partial \tilde{x}^{\nu}}g_{\alpha\beta}(x,y)\, .
\end{equation}
Due 
the 
property \ref{metric-transf},  
$g_{\mu\nu}$ is referred here 
as the components of a {distinguished tensor field} (or $d$-tensor field) on the manifold $\widetilde{TM}$, 
which follows 
the notation adopted 
in~Ref. 
\cite{Miron:1994nvt}. The extremization of the arc-lenght functional \eqref{arclength1} gives the following geodesic equation, 
\begin{equation}
    \frac{d^2x^{\mu}}{d\mu^2}+2G^{\mu}(x,\dot x)=2\frac{dF}{d\mu}\frac{\partial F}{\partial \dot{x}^{\mu}}\, ,
\end{equation}
where $G^{\mu}=G^{\mu}(x,\dot{x})$ are the spray coefficients~\cite{spray-finsler} and are given in terms of the Christoffel 
symbols, $\gamma^{\alpha}_{\mu\nu}$, 
of the metric $g_{\mu\nu}$:
\begin{align}
    &G^{\alpha}(x,\dot{x})=\frac{1}{2}\gamma^{\alpha}_{\mu\nu}(x,\dot{x})\dot{x}^{\mu}\dot{x}^{\nu}\, ,\\
   &\gamma^{\alpha}_{\mu\nu}(x,\dot{x})=\frac{1}{2}g^{\alpha\beta}\left(\frac{\partial g_{\mu\beta}}{\partial x^{\nu}}+\frac{\partial g_{\nu\beta}}{\partial x^{\mu}}-\frac{\partial g_{\mu\nu}}{\partial x^{\beta}}\right)\, .
\end{align}

If we choose the arc-length parametrization, i.e., the one in which $F=1$, we have a sourceless geodesic equation. 
This expression means that the trajectories generated by a MDR of the form $H(x,\dot{x})=m^2$ are, 
 actually,  
geodesics of a Finsler metric. The presence of spray coefficients allows us to construct another 
quite a
useful 
quantity, 
the so-called Cartan {non-linear connection}, given by (in this paper, we interchange the notation $\dot{x}\leftrightarrow y$ freely)
\begin{equation}\label{cartan-connection}
    N^{\mu}{}_{  \nu}(x,y)=\frac{\partial}{\partial y^{\nu}}G^{\mu}(x,y)\, ,
\end{equation}
that transforms according to
\begin{equation}
    \tilde{N}^{\mu}{}_{\nu}=\frac{\partial \tilde{x}^{\mu}}{\partial x^{\alpha}}\frac{\partial x^{\beta}}{\partial \tilde{x}^{\nu}}N^{\alpha}{}_{\beta}-\frac{\partial^2 \tilde{x}^{\mu}}{\partial x^{\alpha}\partial x^{\beta}}\frac{\partial x^{\beta}}{\partial \tilde{x}^{\nu}}y^{\alpha}\, .
\end{equation}

The introduction of this quantity allows us to introduce a useful basis of the tangent space of the tangent bundle at each point. In fact, since according to the coordinate transformation \eqref{coordtrans1} and \eqref{vectrans1}, the usual coordinate basis transforms as
\begin{align}
    &\frac{\partial}{\partial \tilde{x}^{\mu}}=\frac{\partial x^{\nu}}{\partial \tilde{x}^{\mu}}\frac{\partial}{\partial x^{\nu}}+\frac{\partial^2 x^{\nu}}{\partial \tilde{x}^{\mu}\partial\tilde{x}^{\alpha}}\frac{\partial \tilde{x}^{\alpha}}{\partial x^{\beta}}y^{\beta}\frac{\partial}{\partial y^{\nu}}\, ,\\
    &\frac{\partial}{\partial \tilde{y}^{\mu}}=\frac{\partial y^{\nu}}{\partial \tilde{y}^{\mu}}\frac{\partial}{\partial y^{\nu}}\, .
\end{align}

In addition, a non-linear connection allows us to define the following frame:
\begin{align}
    &\frac{\delta}{\delta x^{\mu}}=\delta_{\mu}=\frac{\partial}{\partial x^{\mu}}-N^{\nu}{}_{ \mu}\frac{\partial}{\partial y^{\nu}}\, ,\\
    &\dot{\partial}_{\mu}=\frac{\partial}{\partial y^{\mu}}\, .
\end{align}

Due to the transformation properties of the non-linear connection, this basis transforms as
\begin{align}
    \tilde{\delta}_{\mu}=\frac{\partial x^{\nu}}{\partial \tilde{x}^{\mu}}\delta_{\nu}\, ,\\
    \tilde{\dot{\partial}}_{\mu}=\frac{\partial x^{\nu}}{\partial \tilde{x}^{\mu}}\dot{\partial}_{\nu}\, .
\end{align}

This means that one is 
able to split the tangent space of the tangent bundle into 
horizontal,   
$HTM=\text{span}\{\delta_{\mu}\}$,  
and vertical,  
$VTM=\text{span}\{\dot{\partial}_{\mu}\}$, spaces,  
such that $T\widetilde{TM}=HTM\oplus VTM$ in each point $(x,y)$. Similarly, the same reasoning applies to the cotangent space; i.e., we split $T^*\widetilde{TM}=H^*TM\oplus V^*TM$ spanned as $H^*TM=\text{span}\{dx^{\mu}\}$ and $V^*TM=\text{span}\{\delta y^{\mu}\}$, where
\begin{align}
    \delta y^{\mu}=dy^{\mu}+N^{\mu}{}_{ \nu}dx^{\nu}\, ,
\end{align}
which transforms as
\begin{align}
    d\tilde{x}^{\mu}=\frac{\partial \tilde{x}^{\mu}}{\partial x^{\nu}}dx^{\nu}\, ,\\
    \delta \tilde{y}^{\mu}=\frac{\partial \tilde{x}^{\mu}}{\partial x^{\nu}}\delta y^{\nu}\, .
\end{align}

Such a decomposition of the tangent and cotangent vector spaces implies that 
a vector $X$ and a $1$-form $\omega$ with horizontal and vertical terms can read as 
\begin{align}
X=X^{\mu}\delta_{\mu}+\dot{X}^{\mu}\dot{\partial}_{\mu}=X^H+X^V\, ,\\
\omega=\omega_{\mu}dx^{\mu}+\dot{\omega}_{\mu}\delta y^{\mu}=\omega^H+\omega^V\, .
\end{align}
\par
Endowed with this basis, the metric $ \mathbb{G}(x,y)$ of the {configuration space} is described by the so-called Sasaki-Matsumoto lift of the 
metric $g_{\mu\nu}$: 
\begin{equation}
    \mathbb{G}(x,y)=g_{\mu\nu}(x,y)dx^{\mu}\otimes dx^{\nu}+g_{\mu\nu}(x,y)\delta y^{\mu}\otimes \delta y^{\nu}\, .
\end{equation}
\par
\begin{Definition}\label{def:d-tensor}
A tensor field $T$ of type $(m+n,p+q)$ on the manifold $\widetilde{TM}$ is called a distinguished tensor field (or $d$-tensor field) if it has the property

\begin{align}    
T\left(\overset{1}{\omega},...,\overset{m}{\omega},\overset{1}{\tau},...,\overset{n}{\tau},\underset{1}{X},...,\underset{p}{X},\underset{1}{Y},...,\underset{q}{Y}\right)=T\left(\overset{1}{\omega}{}^H,...,\overset{m}{\omega}{}^H,\overset{1}{\tau}{}^V,...,\overset{n}{\tau}{}^V,\underset{1}{X}{}^H,...,\underset{p}{X}{}^H,\underset{1}{Y}{}^V,...,\underset{q}{Y}{}^V\right).
\end{align}

\end{Definition}

This definition implies that one  
can write 
a $d$-tensor $T$ in the preferred frame as
\begin{align}
    T=T^{\mu_1...\mu_m \nu_1...\nu_n}{}_{\alpha_1...\alpha_p \beta_1...\beta_q}\frac{\delta}{\delta x^{\mu_1}}&\otimes...\otimes \frac{\delta}{\delta x^{\mu_m}}\otimes \frac{\partial}{\partial y^{\nu_{1}}}\otimes...\otimes\frac{\partial}{\partial y^{\nu_n}}\nonumber\\
   &\otimes dx^{\alpha_1}\otimes...\otimes dx^{\alpha_p}\otimes\delta y^{\beta_{1}}\otimes...\otimes \delta y^{\beta_q}\, ,
\end{align}
and that it transforms according to the rule, 
\begin{align}
   &\widetilde{T}^{\mu_1...\mu_m \nu_1...\nu_n}{}_{\alpha_1...\alpha_p \beta_1...\beta_q}\label{t-rule}\\
    &=\frac{\partial \tilde{x}^{\mu_1}}{\partial x^{\epsilon_1}}...\frac{\partial \tilde{x}^{\mu_m}}{\partial x^{\epsilon_m}}\frac{\partial \tilde{x}^{\nu_1}}{\partial x^{\lambda_1}}...\frac{\partial \tilde{x}^{\nu_n}}{\partial x^{\lambda_n}} \frac{\partial x^{\gamma_1}}{\partial \tilde{x}^{\alpha_1}}...\frac{\partial x^{\gamma_p}}{\partial \tilde{x}^{\alpha_p}}\frac{\partial x^{\rho_1}}{\partial \tilde{x}^{\beta_1}}...\frac{\partial x^{\rho_q}}{\partial \tilde{x}^{\beta_q}}T^{\epsilon_1..\epsilon_n \lambda_1...\lambda_m}{}_{\gamma_1...\gamma_p \rho_1...\rho_q}\, .\nonumber
\end{align}
An example of $d$-tensor field is the metric whose components are given by Equation~\eqref{metric-transf}.


\subsection{$N$-Linear Connection}\label{n-linear1}
Given a linear connection, $D$, on the manifold $\widetilde{TM}$, if it preserves the parallelism of the horizontal and vertical spaces, i.e., 
if it can be written as
\begin{equation}
D_{\delta_{\nu}}\delta_{\mu}=L^{\alpha}_{\mu\nu}\delta_{\alpha}\, , \qquad 
D_{\delta_{\nu}}\dot{\partial}_{\alpha}=L^{\mu}_{\alpha\nu}\dot{\partial}_{\mu}\, , 
\label{e35}
\end{equation}
\begin{equation}
D_{\dot{\partial}_{\nu}}\delta_{\mu}=C^{\alpha}_{\mu\nu}\delta_{\alpha}\, , \qquad D_{\dot{\partial}_{\nu}}\dot{\partial}_{\mu}=C^{\alpha}_{\mu\nu}\dot{\partial}_{\alpha}\, ,
\label{e36}
\end{equation}
then 
 is called  
an $N$-linear connection. Let us consider a coordinate change; thus, the 
coefficients (\ref{e35}) and (\ref{e36}) transform as
\begin{align}
    &\tilde{L}^{\alpha}_{\mu\nu}=\frac{\partial\tilde{x}^{\alpha}}{\partial x^{\beta}}\frac{\partial x^{\lambda}}{\partial \tilde{x}^{\mu}}\frac{\partial x^{\epsilon}}{\partial \tilde{x}^{\nu}}L^{\beta}_{\lambda\epsilon} + \frac{\partial^2 x^{\beta}}{\partial \tilde{x}^{\mu}\partial \tilde{x}^{\nu}}\frac{\partial\tilde{x}^{\alpha}}{\partial x^{\beta}}\, ,\\
   &\tilde{C}^{\alpha}_{\mu\nu}=\frac{\partial\tilde{x}^{\alpha}}{\partial x^{\beta}}\frac{\partial x^{\lambda}}{\partial \tilde{x}^{\mu}}\frac{\partial x^{\epsilon}}{\partial \tilde{x}^{\nu}}C^{\beta}_{\lambda\epsilon}\, .
\end{align}

Endowed with these coefficients, the derivative of a $d$-tensor can be decomposed into a horizontal and a vertical parts, 
such that one 
can apply the covariant derivative of a tensor $T$ of type $(m+n,p+q)$ in the direction of a vector $X$ as a
direction of a vector $X$ as 
\begin{align}
    &D_XT=D_{X^H}T+D_{X^V}T\nonumber\\
    &=\left(T^{\mu_1...\mu_m \nu_1...\nu_n}{}_{\alpha_1...\alpha_p \beta_1...\beta_q|\epsilon} X^{\epsilon}+T^{\mu_1...\mu_m \nu_1...\nu_n}{}_{\alpha_1...\alpha_p \beta_1...\beta_q||\epsilon} \dot{X}^{\epsilon}\right)\frac{\delta}{\delta x^{\mu_1}}\otimes...\otimes \frac{\delta}{\delta x^{\mu_m}}\nonumber\\
    &\otimes \frac{\partial}{\partial y^{\nu_1}}\otimes...\otimes\frac{\partial}{\partial y^{\nu_{n}}}\otimes dx^{\alpha_1}\otimes...\otimes 
dx^{\alpha_p}\otimes\delta y^{\beta_1}\otimes...\otimes \delta y^{\beta_q} \, ,
\end{align}
where

\begin{align}
    &T^{\mu_1...\mu_m \nu_1...\nu_n}{}_{\alpha_1...\alpha_p \beta_1...\beta_q|\epsilon}\\
    &=\frac{\delta}{\delta x^{\epsilon}}T^{\mu_1...\mu_m \nu_1...\nu_n}{}_{\alpha_1...\alpha_p \beta_1...\beta_q}+L^{\mu_1}_{\gamma\epsilon}T^{\gamma...\mu_m \nu_1...\nu_n}{}_{\alpha_1...\alpha_p \beta_1...\beta_q}+...-L^{\gamma}_{\alpha_{1}\epsilon}T^{\mu_1...\mu_m \nu_1...\nu_n}{}_{\gamma...\alpha_p \beta_1...\beta_q}\nonumber\, ,\\
    &T^{\mu_1...\mu_m \nu_1...\nu_n}{}_{\alpha_1...\alpha_p \beta_1...\beta_q||\epsilon}\\
    &=\frac{\partial}{\partial y^{\epsilon}}T^{\mu_1...\mu_m \nu_1...\nu_n}{}_{\alpha_1...\alpha_p \beta_1...\beta_q}+C^{\mu_1}_{\gamma\epsilon}T^{\gamma...\mu_m \nu_1...\nu_n}{}_{\alpha_1...\alpha_p \beta_1...\beta_q}+...-C^{\gamma}_{\alpha_1\epsilon}T^{\mu_1...\mu_m \nu_1...\nu_n}{}_{\gamma...\alpha_p \beta_1...\beta_q}\nonumber\, 
 , 
\end{align}
and 
the property that the covariant derivative is linear in the direction $X$ is used.  
The triple $D\Gamma(N,L,C)$ describes the parallel transport and decomposition of the tangent and cotangent spaces of the tangent bundle into horizontal and vertical spaces. At this point, we need to comment on some remarkable $N$-linear connections that are considered in the literature.

The first connection 
is the {metrical Cartan connection,} 
$C\Gamma(N^{\mu}{}_{\nu},L^{\alpha}_{\mu\nu},C^{\alpha}_{\mu\nu})$. In this case, $N^{\mu}{}_{\nu}$ is given by the canonical Cartan non-linear connection, defined by the spray coefficients \eqref{cartan-connection}. The coefficients $L^{\alpha}_{\mu\nu}$ and $C^{\alpha}_{\mu\nu}$ are given, respectively, by
\begin{align}
    L^{\alpha}_{\mu\nu}=\frac{1}{2}g^{\alpha\beta}\left(\frac{\delta g_{\mu\beta}}{\delta x^{\nu}}+\frac{\delta g_{\nu\beta}}{\delta x^{\mu}}-\frac{\delta g_{\mu\nu}}{\delta x^{\beta}}\right)\, ,\label{l-cartan}\\
    C^{\alpha}_{\mu\nu}=\frac{1}{2}g^{\alpha\beta}\left(\frac{\delta g_{\mu\beta}}{\delta y^{\nu}}+\frac{\delta g_{\nu\beta}}{\delta y^{\mu}}-\frac{\delta g_{\mu\nu}}{\delta y^{\beta}}\right)\, .\label{c-cartan}
\end{align}
This connection is metrical (i.e., without non-metricity tensors)  considering both horizontal and vertical covariant derivatives of the Finsler metric.
\par
Besides, 
the {Berwald connection} is given by the triple $B\Gamma(N^{\mu}{}_{\nu}, \partial N^{\alpha}{}_{\mu}/\partial y^{\nu},0)$ and presents horizontal and vertical non-metricities.  
The {Chern--Rund connection,} 
$R\Gamma(N^{\mu}{}_{\nu},L^{\alpha}_{\mu\nu},0)$,  
is horizontally metrical, but represents vertical non-metricity. Additionally, the {Hashiguchi connection,} 
$H\Gamma(N^{\mu}{}_{\nu},\partial N^{\alpha}{}_{\mu}/\partial y^{\nu},C^{\alpha}_{\mu\nu})$,  
represents  
horizontal non-metricity, but it is vertically metrical. In these expressions, $N$ is the canonical Cartan non-linear connection \eqref{cartan-connection}, 
$L$ is given by Equation 
\eqref{l-cartan}, and $C$ is given by Equation 
$\eqref{c-cartan}$.

\subsection{Symmetries}\label{symmetries1}

Geometrical language naturally realizes the concept of symmetry of physical equations. General relativity 
given 
in terms of Riemannian geometry encompasses the invariance under general coordinate transformations, and the isometries of the Minkowski space describe the Poincar\'e transformations (actually, one can further apply this technique for maximally symmetric spaces, including de Sitter and anti-de Sitter ones). Finsler geometry, as we have been using, allows us to go beyond this scope and to define deformed Lorentz/Poincar\'e transformations that present Planck scale corrections even in the presence of a local modified dispersion relation. One can see how this will naturally emerge, since the invariance of the arc-length \eqref{arclength1} is compatible with  the invariance of the action in the Hamiltonian formulation \eqref{action-ham1}, from which such an 
arc-length was derived. This idea was firstly noticed 
in~Ref. 
\cite{Girelli:2006fw} and later explicitly explored in 
Refs.~\cite{Amelino-Camelia:2014rga,Lobo:2016xzq}. The master equation for this purpose is the one that follows from the invariance of the 
Finslerian interval $ds^2$, as done in 
Appendix 
 A 
of~Ref. 
\cite{Amelino-Camelia:2014rga}. From this invariance,  the Finslerian killing equation for the killing vector was found, with components $\xi^{\alpha}$, which should be solved in order to derive the deformed symmetries in the DSR 
context, 
\begin{equation}
    \xi^{\alpha}\partial_{\alpha}g_{\mu\nu}+g_{\alpha\nu}\partial_{\mu}\xi^{\alpha}+g_{\mu\alpha}\partial_{\nu}\xi^{\alpha}+y^{\alpha}\partial_{\alpha}\xi^{\beta}\dot{\partial}_{\beta}g_{\mu\nu}=0\, .
\end{equation}

\subsection{Finsler--q-de Sitter (Tangent Bundle Case)}
As an example that presents a non-trivial non-linear connection, we shall consider the case of a Finsler geometry inspired by the so-called $q$-de 
Sitter deformed relativity. This case has been previously studied in the literature, 
e.g., in~Refs. 
\cite{Barcaroli:2015eqe,Letizia:2016lew,Lobo:2016xzq,Lobo:2016lxm}, and can be described by an algebra that deforms the one of Poincar\'e in a way that gives the de Sitter symmetry when a quantum gravity parameter goes to zero, and on the other hand, gives the so-called $\kappa$-Poincar\'e algebra (that deforms the Poincar\'e one by an energy scale parameter, supposedly the Planck energy) when the de Sitter curvature parameter goes to zero. Therefore, it corresponds to an authentic realization of a deformed relativity scenario, even in the presence of what can be interpreted as spacetime curvature. In this subsection, 
we 
initially consider results that were originally presented 
in~Ref. 
\cite{Letizia:2016lew} in $1+1$ dimensions.
\par
The MDR related to this algebra (in a given basis) can be perturbed to first order in the Planck length and de Sitter curvature parameters $\ell$ and $H$, respectively, as
\begin{equation}\label{q-hamiltonian}
    {\cal H}(x,p)=p_0^2-p_1^2(1+\ell p_0)(1-2H x^0)\, .
\end{equation}

By using the action given by Equation 
\eqref{action-ham1} and the algorithm that follows it, the following Finsler function can be obtained:
\begin{equation}\label{qf-metric1}
 F(x,\dot{x})=\sqrt{(\dot{x}^0)^2-(1-2Hx^0)(\dot{x}^1)^2}+\ell\frac{m}{2}\frac{(1-2Hx^0)\dot{x}^0(\dot{x}^1)^2}{(\dot{x}^0)^2-(1-2Hx^0)(\dot{x}^1)^2}\, ,
\end{equation}
from which the Finsler metric can be found from Equation~\eqref{f-metric1}: 

\begin{eqnarray}
g_{\mu\nu}^F(x,\dot{x})=\begin{pmatrix}
1+\frac{3a^4m\ell\dot{x}^0(\dot{x}^1)^4}{2[(\dot{x}^0)^2-a^2(\dot{x}^1)^2]^{5/2}} & \frac{m\ell a^4(\dot{x}^1)^3[a^2(\dot{x}^1)^2-4(\dot{x}^0)^2]}{2[(\dot{x}^0)^2-a^2(\dot{x}^1)^2]^{5/2}} \\
\frac{m\ell a^4(\dot{x}^1)^3[a^2(\dot{x}^1)^2-4(\dot{x}^0)^2]}{2[(\dot{x}^0)^2-a^2(\dot{x}^1)^2]^{5/2}} & -a^2+\frac{m\ell a^2(\dot{x}^0)^3[2(\dot{x}^0)^2+a^2(\dot{x}^1)^2]}{2[(\dot{x}^0)^2-a^2(\dot{x}^1)^2]^{5/2}} 
\end{pmatrix}\, ,
\end{eqnarray}
where $a=a(t)=e^{Ht}=1+Ht+{\cal O}(H^2)$ (in this paper, 
the 
terms that grow with higher orders of $H$ and $\ell$ are discarded). 
The geodesic equation is found from the extremization of the Finsler 
arc-length  
defined by $F$, from which Christoffel symbols and spray coefficients can be calculated. 
 Actually,  
the $\gamma^{\alpha}_{\mu\nu}(x,\dot{x})$ are given, for an arbitrary parametrization, by the set of 
Equations (44) 
 of 
Ref.~\cite{Letizia:2016lew}, from which the spray coefficients are given by

\begin{align}
   G^0(x,\dot{x})= \frac{1}{8} a^2 H (\dot{x}^1)^2 &\left[4-\frac{\ell m \dot{x}^0}{\left[(\dot{x}^0)^2-a^2 (\dot{x}^1)^2\right]^{7/2}}\left(-28 a^6 (\dot{x}^1)^6+12 a^2 (\dot{x}^0)^4 (\dot{x}^1)^2\right.\right.\nonumber\\
  &\left.\left.+a^2 \left(17 a^2+28\right) (\dot{x}^0)^2 (\dot{x}^1)^4+16 (\dot{x}^0)^6\right)\right]\, ,\\
   G^1(x,\dot{x})=H \dot{x}^0\dot{x}^1+\ell&\left[\frac{a^2 H m (\dot{x}^1)^3 \left(a^6 (\dot{x}^1)^6-6 a^4 (\dot{x}^0)^2 (\dot{x}^1)^4+3 a^2 (\dot{x}^0)^4 (\dot{x}^1)^2-28 (\dot{x}^0)^6\right)}{4 \left((\dot{x}^0)^2-a^2 (\dot{x}^1)^2\right)^{7/2}}\right]\, .
\end{align}

As can be seen, these coefficients are $2$-homogeneous in the velocities, as expected. The Cartan non-linear connection coefficients read:

\begin{align}
N^0{}_0(x,\dot{x})=&\frac{H \ell m (\dot{x}^1)^4 \left(-28 (\dot{x}^1)^6-33 (\dot{x}^1)^4 (\dot{x}^0)^2+240 (\dot{x}^1)^2 (\dot{x}^0)^4+136 (\dot{x}^0)^6\right)}{8 \left((\dot{x}^0)^2-(\dot{x}^1)^2\right)^{9/2}}\, ,\\
N^0{}_1(x,\dot{x})=&H \dot{x}^1-\frac{H \ell m \dot{x}^1 \dot{x}^0 }{8 \left((\dot{x}^0)^2-(\dot{x}^1)^2\right)^{9/2}}\left(28 (\dot{x}^1)^8-179 (\dot{x}^1)^6 (\dot{x}^0)^2+306 (\dot{x}^1)^4 (\dot{x}^0)^4\right.\nonumber\\
&\left.+128 (\dot{x}^1)^2 (\dot{x}^0)^6+32 (\dot{x}^0)^8\right), \nonumber\\
N^1{}_0(x,\dot{x})=&H \dot{x}^1+\frac{H \ell m (\dot{x}^1)^3 \dot{x}^0 \left(5 (\dot{x}^1)^6+18 (\dot{x}^1)^4 (\dot{x}^0)^2+159 (\dot{x}^1)^2 (\dot{x}^0)^4+28 (\dot{x}^0)^6\right)}{4 \left((\dot{x}^0)^2-(\dot{x}^1)^2\right)^{9/2}}\, , \\
N^1{}_1(x,\dot{x})=&H \dot{x}^0-\frac{H \ell m (\dot{x}^1)^2 \left(2 (\dot{x}^1)^8-9 (\dot{x}^1)^6 (\dot{x}^0)^2+36 (\dot{x}^1)^4 (\dot{x}^0)^4+97 (\dot{x}^1)^2 (\dot{x}^0)^6+84 (\dot{x}^0)^8\right)}{4 \left((\dot{x}^0)^2-(\dot{x}^1)^2\right)^{9/2}}\, ,
\end{align}

where the worldlines are autoparallel curves of this non-linear connection. 
Let us note 
that some terms of the connection are only present due to the coupling between the spacetime curvature 
parameter, $H$, 
and the one that gives a non-trivial velocity space,  
$\ell$. Some curvature-triggered effects in quantum gravity have been recently analyzed~\cite{Amelino-Camelia:2020bvx}.
\par
Endowed with these coefficients, the preferred frames that induce the horizontal and vertical decomposition can be immediately found, in addition the $N$-linear connection coefficients $L^{\alpha}_{\mu\nu}$ and $C^{\alpha}_{\mu\nu}$, as discussed in 
Section~\ref{sec:rainbow}. 
 Till now, 
only
kinematical properties were discussed, 
 but the choice of the given connection should be given either by physical conditions imposed on the dynamics of the 
spacetime or by possible effective gravitational field equations for a quantum configuration space.
\par
To finalize this Section, let us 
discuss the symmetries of the spacetime. A deep analysis of the killing vectors of the $H\rightarrow 0$ limit of this Finsler framework was carried out 
in~Ref. 
\cite{Barcaroli:2015eqe}. Even in that 
simplified scenario, the equations are quite lengthy 
 which we omit here. 
However, some properties should be mentioned. Firstly, the transformations generated by the killing 
vectors seem to not exactly preserve the line element, but contribute with a term that is given by a total derivative in the action parameter; therefore, the kinematical results of these two line elements coincide. Secondly, the results found are compatible with the $\kappa$-Poincar\'e scenario that inspired this approach. From the Finsler perspective, it is possible to derive more general results, but they reduces to those of the bicrossproduct basis of $\kappa$-Poincar\'e by an appropriate choice of free functions and parameters. The third point is that a finite version of transformations that preserve the $\kappa$-Poincar\'e dispersion relation was recently made 
in~Ref. 
\cite{Lobo:2021yem} through an alternative approach, which does not rely on the killing vectors but is determined by the Finsler function and 
the definition of momentum 
 (explored in Section~\ref{sec:dualfinsler} below); 
however, a complete integration of the finite isometry and a comparison between these approaches is still missing in the literature. To finalize, the case of $H\neq 0$ was investigated 
in~Ref. 
\cite{Lobo:2016xzq}, but in conformal coordinates (which are not the ones that 
are considered 
in this application), and was not done in so much detail as the flat case, but a generator of the corresponding curved boost transformation was made explicit in 
Equation (25) 
of~Ref. 
\cite{Lobo:2016xzq}.

\section{The Cotangent Bundle Version of Finsler Geometry}\label{sec:dualfinsler}

As was 
discussed in~Ref. 
\cite{Girelli:2006fw}, by mapping the velocity of the particle to its momentum, it is possible to find the version of the Finsler metric defined in the cotangent bundle or phase space. 
Already 
from the definition of the 4-momentum,  
\begin{equation}\label{kin-map1}
    p_{\mu}=m\frac{\partial F}{\partial y^{\mu}}\, ,
\end{equation}
when it is possible to invert this expression to find $y=y(p)$, one 
can substitute this result in the Finsler metric as $h^F_{\mu\nu}(x,p)=g_{\mu\nu}^F(x,y(p))$. This metric is defined on the slit cotangent 
bundle, $\widetilde{T^*M}=T^*M/\{0\}$, 
where we also remove the zero section in each spacetime point for the same technical reasons as 
discussed in Section~\ref{sec:finsler} above. 
Since the 
  quantities are now 
defined in the cotangent bundle, we need to also address some issues that were raised in 
Section~\ref{sec:finsler}
concerning the 
tangent bundle. This Section's 
 notation is applied according 
to~Ref. 
\cite{Miron:1994nvt}. For instance, under a change of coordinates, the spacetime and momentum variables transformed 
according to
    
    \begin{align}\label{coord2}
    \tilde{x}^{\mu}&=\tilde{x}^{\mu}(x)\, ,\\
    \tilde{p}_{\mu}&=\frac{\partial x^{\nu}}{\partial \tilde{x}^{\mu}}p_{\nu}\, ,
    \end{align}
which means that the frame $(\partial/\partial x^{\mu},\partial/\partial p_{\nu})$ transforms as
    \begin{align}
   \frac{\partial}{\partial \tilde{x}^{\mu}}&=\frac{\partial x^{\nu}}{\partial \tilde{x}^{\mu}}\frac{\partial}{\partial x_{\nu}}+\frac{\partial p_{\nu}}{\partial \tilde{x}^{\mu}}\frac{\partial}{\partial p_{\nu}}\, ,\\
   \frac{\partial}{\partial \tilde{p}_{\mu}}&=\frac{\partial \tilde{x}^{\mu}}{\partial x^{\nu}}\frac{\partial}{\partial p_{\nu}}\, .
 \end{align}
    
On the other hand, the natural coframe $(dx^{\mu},dp_{\nu})$ changes as
   \begin{align}
   d\tilde{x}^{\mu}&=\frac{\partial\tilde{x}^{\mu}}{\partial x^{\nu}}dx^{\nu}\, ,\\
   d\tilde{p}_{\mu}&=\frac{\partial x^{\nu}}{\partial \tilde{x}^{\mu}}dp_{\nu}+\frac{\partial^2 x^{\nu}}{\partial \tilde{x}^{\mu}\partial\tilde{x}^{\lambda}}p_{\nu}d\tilde{x}^{\lambda}\, .
   \end{align}

Simlarly to that in Section~\ref{sec:finsler}, 
the presence of a nonlinear connection,   
$O_{\mu\nu}$,  
allows one to split the cotangent bundle into a horizontal and a vertical subbundle. Inspired by the consideration of 
the Hamilton case considered 
in~Ref. 
\cite{Barcaroli:2015xda} 
 (discussed below),  
we propose the following dual non-linear connection (constructed in Appendix \ref{app:conn}):
\begin{equation}\label{dual-conn1}
    O_{\mu\nu}(x,p)=-m\left[N^{\alpha}{}_{\mu}\frac{(g_{\alpha\nu}-p_{\alpha}p_{\nu}/m^2)}{F}-\partial_{\mu}\dot{\partial}_{\nu}F\right]\Bigg{|}_{(x,y(p))}\, ,
\end{equation}
where $p=p(y)$ is the kinematical map defined by Equation~\eqref{kin-map1}. By construction, these symbols have the transformation properties of 
a nonlinear connection,   
\begin{equation}
\tilde{O}_{\mu\nu}=\frac{\partial x^{\lambda}}{\partial \tilde{x}^{\mu}}\frac{\partial x^{\epsilon}}{\partial \tilde{x}^{\nu}}O_{\lambda\epsilon} + \frac{\partial^2 x^{\beta}}{\partial \tilde{x}^{\mu}\partial \tilde{x}^{\nu}}p_{\beta}\, .
\end{equation}

Endowed with a nonlinear connection $O_{\mu\nu}$, 
one  
can decompose the tangent bundle of the cotangent bundle by the Whitney sum in each point $T_u\widetilde{T^*M}=O_u\oplus V_u,\,   \forall u\in \widetilde{T^*M}$. 
 The subbundle $O_u$ is called horizontal space and is spanned by the frame,  
\begin{equation}
\frac{\delta}{\delta x^{\mu}}=\delta_{\mu}=\frac{\partial}{\partial x^{\mu}}+O_{\mu\nu}\frac{\partial}{\partial p_{\nu}}\, ,
\end{equation}
and the subbundle $V_u$ is called vertical space and is spanned by the frame in each point of $\widetilde{T^*M}$: 
\begin{equation}
\bar{\partial}^{\mu}=\frac{\partial}{\partial p_{\mu}}\, ,
\end{equation}
such that $T_u\widetilde{T^*M}=\text{span}\{\delta_{\mu},\bar{\partial}^{\nu}\}$. The transformation properties of the nonlinear connection are 
implied in the following rule for transforming this basis:
\begin{align}
\frac{\delta}{\delta \tilde{x}^{\mu}}=\tilde{\delta}_{\mu}=\frac{\partial x^{\nu}}{\partial \tilde{x}^{\mu}}\frac{\delta}{\delta x^{\nu}}=\frac{\partial x^{\nu}}{\partial \tilde{x}^{\mu}}\delta_{\nu}\, ,\label{cot-transf11}\\
\frac{\partial}{\partial \tilde{p}_{\mu}}=\tilde{\bar{\partial}}^{\mu}=\frac{\partial \tilde{x}^{\mu}}{\partial x^{\nu}}\frac{\partial}{\partial p_{\nu}}=\frac{\partial \tilde{x}^{\mu}}{\partial x^{\nu}}\bar{\partial}^{\nu}\, .\label{cot-transf12}
\end{align}

Equivalently, with the nonlinear connection, we can decompose the cotangent space $T^*_u\widetilde{T^*M}=\text{span}\{dx^{\mu},\delta p_{\nu}\}$, where
\begin{equation}
\delta p_{\mu}=dp_{\mu}-O_{\nu\mu}dx^{\nu}\, .
\end{equation}
Therefore, the dual basis transforms as
\begin{align}
d\tilde{x}^{\mu}&=\frac{\partial \tilde{x}^{\mu}}{\partial x^{\nu}}dx^{\nu}\, ,\label{cot-transf21}\\
\delta \tilde{p}_{\mu}&=\frac{\partial x^{\nu}}{\partial \tilde{x}^{\mu}}\delta p_{\nu}\, . \label{cot-transf22}
\end{align}

Similarly to what has been done for the tangent bundle case, such a decomposition allows us to 
express 
a vector and a $1$-form 
 via  
horizontal and vertical components, where now, the vertical component is considered 
along momenta instead of velocities,
\begin{align}
X=X^{\mu}\delta_{\mu}+\bar{X}_{\mu}\bar{\partial}^{\mu}=X^H+X^V\, ,\\
\omega=\omega_{\mu}dx^{\mu}+\bar{\omega}^{\mu}\delta p_{\mu}=\omega^H+\omega^V\, .
\end{align}

Besides, 
the metric $\mathbb{H}(x,p)$ of the {configuration space} is defined as follows. Given a metric $h^{\mu\nu}(x,p)$, and the 
nonlinear connection $O_{\mu\nu}(x,p)$, the quantum phase space presents metrical properties given by the tensor, 
\begin{equation}
\mathbb{H}(x,p)=h_{\mu\nu}(x,p)dx^{\mu}\otimes dx^{\nu}+h^{\mu\nu}(x,p)\delta p_{\mu}\otimes \delta p_{\nu}\, .
\end{equation}

We refer to the tensor $\mathbb{H}$ as the $N$-lift to $\widetilde{T^*M}$ of the metric $h_{\mu\nu}$. The map between $y$ and $p$ cannot be done, in general, involving quantities that are parametrization-dependent because $p$ itself is parametrization-invariant, whereas $y$ is not. That is why 
one can only 
assume $y(p)$ for the definition of the metric $h^F_{\mu\nu}$.
\par
Endowed with these quantities, 
one 
can 
just 
extend the definition of $d$-tensors \ref{def:d-tensor} to the cotangent case, in which 
 one only needs  
to consider the use of the nonlinear connection $O_{\mu\nu}$ and the adapted basis defined in this Section. 

 The above  
implies that  
a  $d$-tensor $T$ of type $(m+q,n+p)$ can be rewritten 
in the preferred basis as
\begin{align}
    T=T^{\mu_1...\mu_m}{}_{ \nu_1...\nu_n\alpha_1...\alpha_p}{}^{\beta_1...\beta_q}&\frac{\delta}{\delta x^{\mu_1}}\otimes...\otimes \frac{\delta}{\delta x^{\mu_m}}\otimes \frac{\partial}{\partial p_{\nu_{1}}}\otimes...\otimes\frac{\partial}{\partial p_{\nu_n}}\nonumber\\
   &\otimes dx^{\alpha_1}\otimes...\otimes dx^{\alpha_p}\otimes\delta p_{\beta_{1}}\otimes...\otimes \delta p_{\beta_q}\, ,
\end{align}
whose components transform according to usual linear transformation rules, as the one of Equation 
\eqref{t-rule}.


\subsection{N-Linear Connection}\label{n-linear2}

Equivalently, the notion of differentiation can be defined in the cotangent bundle through the $N$-linear connection $D$, which has the following 
coefficients in the frame $(\delta_{\mu},\bar{\partial}^{\nu})$ (see Theorem 4.9.1 
in~Ref. 
\cite{Miron:1994nvt}):
\begin{align}
D_{\delta_{\nu}}\delta_{\mu}=H^{\alpha}_{\mu\nu}\delta_{\alpha}\, , \qquad D_{\delta_{\nu}}\bar{\partial}^{\mu}=-H^{\mu}_{\alpha\nu}\bar{\partial}^{\alpha}\, ,\\
D_{\bar{\partial}^{\nu}}\delta_{\mu}=C^{\alpha\nu}_{\mu}\delta_{\alpha}\, , \qquad D_{\bar{\partial}^{\nu}}\bar{\partial}^{\mu}=-C_{\alpha}^{\mu\nu}\bar{\partial}^{\alpha}\, .
\end{align}
Otherwise, in the frame $(dx^{\mu},\delta p_{\nu})$ one has 
(see Proposition 4.9.1 
in~Ref. 
\cite{Miron:1994nvt})
\begin{align}
D_{\delta_{\nu}}dx^{\mu}=-H^{\mu}_{\alpha\nu}dx^{\alpha}\, , \qquad D_{\delta_{\nu}}\delta p_{\mu}=H^{\alpha}_{\mu\nu}\delta p_{\alpha}\, ,\\
D_{\bar{\partial}^{\nu}}dx^{\mu}=-C^{\mu\nu}_{\alpha}dx^{\alpha}\, , \qquad D_{\bar{\partial}^{\nu}}\delta p_{\mu}=C_{\mu}^{\alpha\nu}\delta p_{\alpha}\, .
\end{align}

Considering a $N$-linear connection $D$ with set of coefficients,   
$D\Gamma(N)=(H^{\alpha}_{\mu\nu},C^{\alpha}_{\mu\nu})$, one 
can add to it a nonlinear connection, $N_{\mu\nu}$,  
that is in general 
independent of the coefficients of $D$, such that 
 the  
new set is $D\Gamma=(N_{\mu\nu},H^{\alpha}_{\mu\nu},C^{\alpha}_{\mu\nu})$. For this reason, the derivative of a $d$-tensor in the cotangent bundle presents similar usual rules for dealing with up and down indices:

\begin{align}
    &T^{\mu_1...\mu_m}{}_{ \nu_1...\nu_n\alpha_1...\alpha_p}{}^{ \beta_1...\beta_q}{}_{|\epsilon}\\
    &=\frac{\delta}{\delta x^{\epsilon}}T^{\mu_1...\mu_m}{}_{ \nu_1...\nu_n\alpha_1...\alpha_p}{}^{ \beta_1...\beta_q}+H^{\mu_1}_{\gamma\epsilon}T^{\gamma...\mu_m}{}_{ \nu_1...\nu_n\alpha_1...\alpha_p}{}^{ \beta_1...\beta_q}+...-H^{\gamma}_{\nu_1\epsilon}T^{\mu_1...\mu_m}{}_{ \gamma...\nu_n\alpha_1...\alpha_p}{}^{ \beta_1...\beta_q}\nonumber\, ,\\
    &T^{\mu_1...\mu_m}{}_{ \nu_1...\nu_n\alpha_1...\alpha_p}{}^{ \beta_1...\beta_q}{}^{||\epsilon}\\
    &=\frac{\partial}{\partial p_{\epsilon}}T^{\mu_1...\mu_m}{}_{ \nu_1...\nu_n\alpha_1...\alpha_p}{}^{ \beta_1...\beta_q}{}+C^{\mu_1\epsilon}_{\gamma}T^{\gamma...\mu_m}{}_{ \nu_1...\nu_n\alpha_1...\alpha_p}{}^{ \beta_1...\beta_q}{}+...-C^{\gamma\epsilon}_{\nu_1}T^{\mu_1...\mu_m}{}_{ \gamma...\nu_n\alpha_1...\alpha_p}{}^{ \beta_1...\beta_q}{}\nonumber\, .
\end{align}

 Let us note  
that from the kinematical map relating velocities and momenta, the coefficients $H^{\alpha}_{\mu\nu}(x,y(p))$ and 
$C^{\alpha}_{\mu\nu}(x,y(p))$ can be found 
 as been 
parametrization-invariant.


\subsection{Finsler--q-de Sitter (Cotangent Bundle Case)}

Here, we 
again consider the $q$-de Sitter-inspired case. 
Then, 
using  
the Finsler function 
\eqref{qf-metric1}, 
the momentum is 
given 
by Equation:  
\eqref{kin-map1}
\begin{align}
   &p_0=\frac{m\dot{x}^0}{\sqrt{(\dot{x}^0)^2-a^2(\dot{x}^1)^2}}-\ell\frac{m^2a^2(\dot{x}^1)^2(a^2(\dot{x}^1)^2+(\dot{x}^0)^2)}{2[(\dot{x}^0)^2-a^2(\dot{x}^1)^2)]^2}\, ,\\
    &p_1=-\frac{ma^2\dot{x}^1}{\sqrt{(\dot{x}^0)^2-a^2(\dot{x}^1)^2}}+\ell\frac{m^2a^2(\dot{x}^0)^3\dot{x}^1}{((\dot{x}^0)^2-a^2(\dot{x}^1)^2))^2}\, ,
\end{align}
which furnishes a 
 helpful 
expression that is 
throughout this Section 
and is a common trick when trying to find momentum-dependent quantities from the 
 Finsler approach:
\begin{align}
   & \frac{m\dot{x}^0}{\sqrt{(\dot{x}^0)^2-a^2(\dot{x}^1)^2}}=p_0+\ell\frac{a^{-2}(p_1)^2(a^{-2}(p_1)^2+(p_0)^2)}{2m^2}\, ,\\
   &  \frac{ma\dot{x}^1}{\sqrt{(\dot{x}^0)^2-a^2(\dot{x}^1)^2}}=-a^{-1}p_1\left(1+\ell \frac{(p_0)^3}{m^2}\right)\, .
\end{align}

The above expressions allow us to express the Finsler metric through its momentum dependence:

\begin{eqnarray}\label{f-rainb}
g_{\mu\nu}^F(x,\dot{x}(p))=h_{\mu\nu}^F(x,p)=\begin{pmatrix}
1+\frac{3\ell p_0(p_1)^4}{m^4} & -\frac{\ell  a(p_1)^3[(p_1)^2-4(p_0)^2]}{2m^4} \\
-\frac{\ell  a(p_1)^3[(p_1)^2-4(p_0)^2]}{2m^4} & -a^2+\frac{\ell a^2(p_0)^3[2(p_0)^2+(p_1)^2]}{m^4} 
\end{pmatrix}\, ,
\end{eqnarray}
which can be called 
a ''Finsler-rainbow metric.''  

One  
can also find the induced non-linear connection in the cotangent bundle through the 
definition 
 \eqref{dual-conn1} to read as

\begin{align}\label{fcot-conn}
   O_{00}(x,p)=&-\frac{H \ell (p_1)^2}{8 m^{10}}\left[4 (p_0)^{10}+44 (p_0)^8 (p_1)^2+190 (p_0)^6 (p_1)^4-196 (p_0)^4 (p_1)^6\right.\nonumber\\
   &\left.+31 (p_0)^2 (p_1)^8+32 (p_1)^{10}\right]\, ,\\
    O_{01}(x,p)=&H p_1-\frac{\ell H p_0p_1}{8m^{10}}\left[-4 m^8 (p_0)^2+8 (p_0)^{10}+32 (p_0)^8 (p_1)^2+206 (p_0)^6 (p_1)^4\right.\nonumber\\
    &\left.-212 (p_0)^4 (p_1)^6+43 (p_0)^2 (p_1)^8+28 (p_1)^{10}\right]\, ,\\
    O_{10}(x,p)=&H p_1-\frac{H \ell p_0 p_1 }{8m^{10}}\left(-4 m^8 (p_0)^2+4 (p_0)^{10}+140 (p_0)^8 (p_1)^2+2 (p_0)^6 (p_1)^4\right.\nonumber\\
    &\left.-106 (p_0)^4 (p_1)^6+61 (p_0)^2 (p_1)^8+4 (p_1)^{10}\right)\, ,\\
    O_{11}(x,p)=&H p_0+\frac{H \ell}{8 m^{10}}\left(4 (p_0)^2 (p_1)^2 \left(m^8+3 (p_1)^8\right)+8 m^8 (p_1)^4-8 (p_0)^{12}-124 (p_0)^{10} (p_1)^2\right.\nonumber\\
   &\left.-30 (p_0)^8 (p_1)^4+138 (p_0)^6 (p_1)^6-89 (p_0)^4 (p_1)^8-4 (p_1)^{12}\right)\, .
\end{align}

From these expressions, one 
can construct the decomposition of the tangent and cotangent spaces of the cotangent bundle into horizontal and vertical parts, accordingly.

\section{Geometry of the Cotangent Bundle: Hamilton Geometry}\label{sec:hamilton}

Besides the 
Finsler geometry, another interesting proposal for building a natural geometry for propagation of particles that probe a modified dispersion relation consists of the so-called Hamilton geometry. In this case, 
 different 
from the Finsler geometry, 
we 
start 
with a geometric structure defined in the cotangent bundle (the definitions used in this metric follow 
that in 
the book~\cite{Miron:1994nvt} and in 
papers~\cite{Barcaroli:2015xda,Barcaroli:2016yrl,Barcaroli:2017gvg,Pfeifer:2018pty}). 

A Hamilton space is a pair,   
$(M,H(x,p))$, where $M$ is a smooth manifold and $H:T^*M\rightarrow {\mathbb R}$ is a continuous function on the cotangent bundle that satisfies the 
following properties: 
\begin{enumerate}
    \item $H$ is smooth on the manifold $\widetilde{T^*M}$; 
    \item the  
 Hamilton metric, $h_H$, with components, 
    \begin{equation}\label{h-metric1}
        h_H^{\mu\nu}(x,p)=\frac{1}{2}\frac{\partial}{\partial p_{\mu}}\frac{\partial}{\partial p_{\nu}}H(x,p)\, ,
    \end{equation}
    is nondegenerate.
\end{enumerate}

Since one does 
 not have an arc-length functional, worldlines as extremizing curves are an absent concept in this approach. Instead, the equations of motion of a particle that obeys a given Hamiltonian are given by the Hamilton equations of motion:
\begin{align}\label{ham-eqs1}
    \dot{x}^{\mu}&=\frac{\partial H}{\partial p_{\mu}}\, ,\\
    \dot{p}_{\mu}&=-\frac{\partial H}{\partial \dot{x}^{\mu}}\, .
\end{align}

Since 
 this is  
just another metric structure defined in the cotangent bundle, the same results regarding the tools for coordinate transformations given by 
Equation~\eqref{coord2} are applicable here. 
As the case of Hamiltonian mechanics, the definition of Poisson brackets is 
 useful enough 
for our purposes. 
For two real valued functions $F(x,p)$ and $G(x,p)$, their Poisson brackets are given 
in~\cite{Barcaroli:2015xda} 
({{the geometry of the cotangent bundle with deformed Hamiltonian can also be described with the language of symplectic geometry, which 
is 
reviewed 
in~Ref. 
\cite{Nozari:2014qja}}}):
\begin{equation}\label{poisson1}
    \{F(x,p),G(x,p)\}=\partial_{\mu}F\bar{\partial}^{\mu}G-\partial_{\mu}G\bar{\partial}^{\mu}F\, .
\end{equation}

As above, 
in order to divide the tangent and cotangent spaces of the cotangent bundle into horizontal and vertical spaces, a non-linear connection is necessary, and the canonical choice is given in Theorem 5.5.1 
of~Ref. 
\cite{Miron:1994nvt} and Definition 2 
of~Ref. 
\cite{Barcaroli:2015xda} as
\begin{equation}\label{conn-h1}
    O_{\mu\nu}(x,p)=\frac{1}{4}(\{h_{\mu\nu}^H,H\}+h^H_{\mu\alpha}\partial_{\nu}\bar{\partial}^{\alpha}H+h^H_{\nu\alpha}\partial_{\mu}\bar{\partial}^{\alpha}H)\, ,
\end{equation}
where $h_{\mu\nu}^H$ is the inverse of the metric $h^{\mu\nu}_H$. This non-linear connection allows us to use the basis $\delta_{\mu}=\partial_{\mu}+O_{\mu\nu}\bar{\partial}^{\nu}$ and $\bar{\partial}^{\mu}$ as a special basis of $T_{(x,p)}\widetilde{T^*M}$, and to use the basis $dx^{\mu}$ and $\delta p_{\mu}=dp_{\mu}-O_{\nu\mu}dx^{\nu}$ as a special basis of $T^*_{(x,p)}\widetilde{T^*M}$, which transforms 
according to Equations  \eqref{cot-transf11}, \eqref{cot-transf12} and \eqref{cot-transf21}, \eqref{cot-transf22}.

Endowed with these coefficients, following 
Theorem 
 5.6.1 
of~Ref. 
\cite{Miron:1994nvt}, there exists a unique $N$-linear connection $D\Gamma(O)=(H^{\alpha}_{\mu\nu},C_{\alpha}^{\mu\nu})$ such that:
\begin{enumerate}
    \item $O_{\mu\nu}$ is the canonical non-linear connection; 
    \item the 
metric $h^{\mu\nu}_H$ is $h$-covariant constant (no horizontal non-metricity):
    \begin{equation}
        D_{\delta_{\alpha}}h_H^{\mu\nu}=0\, ; 
    \end{equation}
      \item the 
metric $h^{\mu\nu}_H$ is $v$-covariant constant (no vertical non-metricity):
      \begin{equation}
        D{\bar{\partial}^{\alpha}}h_H^{\mu\nu}=0\, ; 
    \end{equation}
      \item $D\Gamma(N)$ is horizontally torsion free:
      \begin{equation}
          T^{\alpha}_{\ \mu\nu}=H^{\alpha}_{\mu\nu}-H^{\alpha}_{\nu\mu}=0\, ; 
      \end{equation}
       \item $D\Gamma(N)$ is vertically torsion free:
      \begin{equation}
          S_{\alpha}^{\ \mu\nu}=C_{\alpha}^{\mu\nu}-C_{\alpha}^{\nu\mu}=0\, ; 
      \end{equation}
      \item the 
  triple $(O_{\mu\nu},H^{\alpha}_{\mu\nu},C_{\alpha}^{\mu\nu})$ has coefficients given by
      \begin{align}
          &O_{\mu\nu}(x,p)=\frac{1}{4}(\{h_{\mu\nu}^H,H\}+h^H_{\mu\alpha}\partial_{\nu}\bar{\partial}^{\alpha}H+h^H_{\nu\alpha}\partial_{\mu}\bar{\partial}^{\alpha}H)\, ,\\
          &H_{\alpha}^{\mu\nu}=\frac{1}{2}h_H^{\alpha\beta}(\delta_{\mu}h^H_{\beta\nu}+\delta_{\nu}h^H_{\beta\mu}-\delta_{\beta}h^H_{\mu\nu})\, ,\\
          &C_{\alpha}^{\mu\nu}=-\frac{1}{2}h_{\alpha\beta}^H\bar{\partial}^{\mu}h_H^{\beta\nu}\, .
      \end{align}
\end{enumerate}
This is called a Cartan $N$-linear covariant derivative. Equivalently, the notion of $d$-tensors and their derivatives discussed in Section~\ref{n-linear2} are applicable.


\subsection{Symmetries}

Hamilton geometry also allows one to encompass a DSR language, as was the case for Finsler geometry discussed in Section~\ref{symmetries1}. However, its realization does not come from the invariance of an interval $ds^2$, since 
one does 
not have it, but from the invariance of the Hamiltonian function $H(x,p)$. The approach, which we highlight here, was done starting from Definition 4 of 
 Section 
II-D 
of~Ref. 
\cite{Barcaroli:2015xda}. In a Hamilton space $(M,H)$ with manifold $M$,  
and Hamiltonian $H$, let $X=\xi^{\mu}\partial_{\mu}$ be a vector field in the basis manifold $M$ and $X^C=\xi^{\mu}\partial_{\mu}-p_{\nu}\partial_{\mu}\xi^{\nu}\bar{\partial}^{\mu}$ be the so-called complete lift of $X$ to $\widetilde{T^*M}$. A symmetry of the Hamiltonian is a transformation generated by $X^C$, whose components satisfy
\begin{equation}
    X^C(H)\xi^{\mu}\partial_{\mu}H-p_{\nu}\partial_{\mu}\xi^{\nu}\bar{\partial}^{\mu}H=0\, .
\end{equation}

If one derivates this expression twice with respect to momenta, 
one gets  
the following result:
\begin{equation}
    0=\frac{1}{2}\bar{\partial}^{\mu}\bar{\partial}^{\nu}X^C(H)=\xi^{\alpha}\partial_{\alpha}h_H^{\mu\nu}-h_H^{\mu\alpha}\partial_{\alpha}\xi^{\nu}-h_H^{\nu\alpha}\partial_{\alpha}\xi^{\mu}-p_{\beta}\partial_{\alpha}\xi^{\beta}\bar{\partial}^{\alpha}h_H^{\mu\nu}\, .
\end{equation}

This is just the generalization of the killing equation to a general Hamilton space. 
 In general,    
if $h_H$ does not depend on momenta, then it reduces to the standard Riemannian case. 
Besides, 
from the expression of the Poisson brackets \eqref{poisson1}, it can verified that such symmetries give rise to conserved charges 
$\xi^{\mu}p_{\mu}$; i.e., that Poisson commutes with the Hamiltonian: 
\begin{equation}
    \{\xi^{\mu}p_{\mu},H\}=0.
\end{equation}
These are the charges that, at an algebraic level, can generate translations, boosts, and rotations, for instance.


\subsection{Hamilton--q-de Sitter (Cotangent Bundle Case)}

As an example, we rely on the results presented 
in~Ref. 
\cite{Barcaroli:2015xda}, which are 
as well  
inspired by the $q$-de Sitter Hamiltonian \eqref{q-hamiltonian}. In this case, the Hamilton metric, 
defined by Equation 
\eqref{h-metric1}, reads: 
\begin{eqnarray}
h^{\mu\nu}_H(x,p)=\begin{pmatrix}
1 & -\ell p_1(1+2Hx^0) \\
-\ell p_1(1+2Hx^0) & -(1+2Hx^0)(1+\ell p_0) 
\end{pmatrix}\, ,
\end{eqnarray}
which, as can be seen, acquires a shape much simpler than the rainbow-Finsler one \eqref{f-rainb} due to the much direct way in which it is calculated. 

The non-linear connection can be read from Equation 
\eqref{conn-h1} and can be cast in a matrix form due to its simplicity:
\begin{eqnarray}\label{omn2}
O_{\mu\nu}(x,p)=\begin{pmatrix}
H\ell p_1^2 & H p_1 \\
H p_1 & H p_0(1-\ell p_0) 
\end{pmatrix}\, .
\end{eqnarray}
As expected, it coincides with the case \eqref{fcot-conn} in the Riemannian case, i.e., when $\ell=0$. 

The Hamilton equations of motion can be found from 
Equation 
\eqref{ham-eqs1} and read: 
\begin{align}
    &\dot{x}^0-2p_0+\ell p_1^2(1+2H x^0)=0\, ,\\
    &\dot{x}^1+2p_1(1+Hx^0)+2\ell p_0p_1(1+2Hx^0)=0\, ,\\
    &\dot{p}_0-2Hp_1^2-2H\ell p_0p_1^2=0\, ,\\
    &\dot{p}_1=0\, .
\end{align}

The autoparallel (horizontal) curves of the non-linear connection satisfy (see Equation 
(8.2) 
in~Ref. 
\cite{Miron:1994nvt})
\begin{equation}
    \dot{p}_{\mu}-O_{\nu\mu}\dot{x}^{\nu}=0\, ,
\end{equation}
and, as can be seen from 
Equation~\eqref{omn2} 
for $O_{\mu\nu}$,  
the worldlines, defined from the Hamilton equations of motion, 
{are not} autoparallels of the non-linear connection.

The symmetries have also been analyzed in~Ref. 
\cite{Barcaroli:2015xda}, where it has been noticed that the conserved charges that generate translations and the boost coincide with 
the 
results from Ref.~\cite{Barcaroli:2015eqe} 
 that do not rely on the geometrical approach 
used in this paper.


\section{The Tangent-Bundle Version of Hamilton Geometry}\label{sec:dualhamilton}

Endowed with Hamilton equations of motion \eqref{ham-eqs1}, one has a map between the momenta and velocities from $\dot{x}^{\mu}=y^{\mu}=\partial H/\partial p_{\mu}$. When it is possible to invert this map to find $p_{\mu}=p_{\mu}(y)$ (as done in 
 Appendix 
 B 
 of~Ref. 
\cite{Barcaroli:2017gvg}), 
one derives  
an interesting map between the cotangent and tangent space version of Hamilton geometry. 
 Indeed,  
using this map, 
a Hamilton metric defined in the tangent bundle reads: 
\begin{align}
    g_H^{\mu\nu}(x,y)\doteq h_H^{\mu\nu}(x,p(y))\, .
\end{align}

The dual non-linear connection in this case has been discussed in 
Appendix C 
 of~Ref. \cite{Barcaroli:2015xda}, and is given by
\begin{equation}
    N(x,y)^{\mu}{}_{\nu}=2O(x,p(y))_{\nu\alpha}h_H^{\alpha\mu}(x,p(y))-(\partial_{\nu}\bar{\partial}^{\mu}H)|_{p=p(y)}\, .
\end{equation}

Its main property is the preservation of the horizontal tangent spaces of the cotangent and tangent bundle connected through the kinematical map $y^{\mu}=\partial H/\partial p_{\mu}$.

With this map, it is possible to define the dual non-linear and $N$-linear connections, now defined in the tangent bundle. It should be stressed that although this gives geometrical quantities defined in the tangent bundle, this does not represent a Finsler geometry, 
 since 
 there is no 
arc-length functional and the Hamilton metric is not, in general, 0-homogeneous to 
start  
with.

\subsection*{Hamilton-$\kappa$-Poincar\'e (Tangent Bundle Case)}
The kinematical map that allows us to describe $y=y(p)$ is found by inverting the relation $y^{\mu}=\partial H/\partial p_{\mu}$ for the $q$-de Sitter Hamiltonian, given by
\begin{align}
    &p_0=\frac{y^0}{2}+\ell\frac{(y^1)^2}{8}\, ,\\
    &p_1=-\frac{y^1}{2}+H\frac{x^0 y^1}{2}+\ell\frac{y^0 y^1}{4}\, .
\end{align}

The metric in the tangent bundle reads: 
\begin{eqnarray}
g^{\mu\nu}_H(x,y)=\begin{pmatrix}
1 & \ell(H  x^0 y^1+y^1)/2 \\
\ell(H  x^0 y^1+y^1)/2 & -(1+2Hx^0) (1+\ell y^0/2) 
\end{pmatrix}\, .
\end{eqnarray}

The dual non-linear connection reads
\begin{eqnarray}
N^{\mu}{}_{\nu}(x,y)=\begin{pmatrix}
-H\ell (y^1)^2/2 & \ell h y^0 y^1+h y^1 \\
H y^1 & -h y^0- 3\ell h (y^1)^2/4 
\end{pmatrix}\, .
\end{eqnarray}

In 
Section~\ref{sec:advdiff} below, 
some key points  of each approach are discussed 
while
comparing the descriptions of configuration and phase 
spaces.

\section{Advantages and Difficulties of Each Formalism}\label{sec:advdiff}

 The approaches considered---Finsler and Hamilton spaces---present 
the  
points that can be considered 
positive or 
negative. 
In this Section, 
we highlight some of 
 those points  
which 
 look to be most important 
from theoretical and phenomenological points of view.
\subsection{Finsler Geometry}

Let us 
emphasize that here 
 not 
a complete 
list  
of positive or negative points are given,
 and, certainly, 
 the points listed 
represent just our view on the 
subject under scrutiny 
 and 
some points 
we are classifying in one way or another 
 can be  
seen by others 
completely differently. 

\subsubsection{Advantages}

{\bf Preservation of the equivalence principle}.  
Due to the presence of an arc-length functional, the extremizing geodesics of the Finsler function are the same worldlines of the 
Hamiltonian,  from which the 
arc-length  
was derived. This means that,  
in the Finslerian language, the equivalence principle is satisfied, 
as soon as 
the worldlines are trajectories of free particles in this spacetime. There is a fundamental difference in comparison to the special or general relativity formulation, since these trajectories are now mass-dependent, since the Finsler function and the metric carry the mass of the particle due to Planck-scale effects. Intriguingly, although the metric does not present a massless limit 
(which 
is  
discussed below), 
it is possible to find trajectories of massless particles, which are compatible with the Hamiltonian formulation, by taking the limit $m\rightarrow 0$ in the geodesic Equation \cite{Amelino-Camelia:2014rga,Lobo:2016xzq}. This 
 finding 
leads to some effects due to modifications of the trajectories of particles. For instance, one of the most explored avenues of quantum gravity phenomenology (maybe competing with threshold effects) is the time delay until particles with different energies might arrive at a detector after a (almost) simultaneous emission~\cite{Jacob:2008bw,Zhu:2022blp} 
(for reviews, 
see~\cite{Amelino-Camelia:2008aez,Addazi:2021xuf}). 
This kind of experimental investigation is not exhausted, and novelties have arrived in the analysis of sets of gamma-ray bursts and candidate neutrinos emitted from them in the multimessenger astronomy approach~\cite{Amelino-Camelia:2016ohi,Amelino-Camelia:2022pja}.
\par
{\bf Preservation of the relativity principle}.   
This formalism allows one to derive and solve the killing equation, which furnishes infinitesimal symmetry transformations of the metric. It has been shown 
in~Ref. 
\cite{Amelino-Camelia:2014rga} that generators of these transformations can be constructed and identified with 
the transformations 
that are generally 
depicted in the doubly special relativity. 
 The latter  
implies, in a preservation of the relativity principle  
that inertial frames should assign the same MDR to a given particle 
 which, in its turn,   
implies that the deformation scale of 
quantum gravity is observer-independent, 
 i.e., two observers would not assign different values, 
in the same system of units, to the quantum gravity scale. 
This preservation has important phenomenological consequences, such as the point 
that the threshold  constraints on the quantum gravity parameter do not apply in the DSR scenario. The reason  
 is that, accompanied by the deformation of the Lorentz (Poincar\'e) symmetries, comes a deformation of the composition law of momenta of particles (for instance $p$ and $q$), such that the nature of interaction vertices to not get modified when transforming from one frame to another:
\begin{equation}
    \Lambda (p\oplus q)=\Lambda(p)\oplus\Lambda(q)\, ,
\end{equation}
where $\Lambda$ is a deformed Lorentz transformation induced by the killing vectors and $\oplus$ represents a modified composition of components of the involved momenta (this covariance condition usually needs a back-reaction on the boost parameter, but we 
do 
not dwell on that here; for more details, 
see~\cite{Majid:2006xn,Lobo:2021yem} and references therein). Threshold constraints, such as the one placed 
in~Ref. 
\cite{HAWC:2019gui}, assumes that the composition of momenta is undeformed, although the dispersion relation is modified in a Lorentz invariance violation (LIV) scenario. When this is the case, processes that are forbidden in special relativity, such as the decay of the photon into an electron--positron pair, becomes kinematicaly allowed for a given threshold energy. The 
no observation of 
 such decays 
allows one to place constraints on the quantum gravity parameter. When the dispersion relation is modified as well, what happens is that 
generally 
these kinds of processes remain forbidden or modifications in the threshold energies are so minute that they are unobservable for a quantum gravity parameter in the order of the Planck energy~\cite{Lobo:2021yem}. This is an important feature of "deforming" instead of "violating" the Lorentz symmetry.
\par
{\bf Preservation of the clock postulate}. 
 The 
 availability of 
an arc-length functional leads 
to a possibility to 
analyze the consequences of having the proper time of a given particle given by it. 
If 
this is the case, then the worldlines or geodesics are just paths that extremize the proper time an observer measures in spacetime, 
similar to that 
in special relativity. One of the consequences of this feature consists of the possibility of connecting the time elapsed in the comoving frame of a particle during its lifetime (which is its lifetime at rest) and the coordinate time, which is the one that is assigned to this phenomenon in the laboratory coordinates. Using this expression, one can investigate the relativistic time dilation (responsible for the "twin paradox") or the so-called first clock effect (for further details on the first and also on the second clock effect, 
which can appear in theories with a non-metricity tensor; see Ref.
~\cite{Lobo:2018zrz}), in which, for instance, the lifetime of a particle is dilated in comparison to the one assigned in the laboratory. Due to Finslerian corrections, the lifetime of a particle in the 
laboratory  
would receive Planckian corrections, which, actually, 
is a novel avenue of phenomenological investigation that is being currently carried out~\cite{Lobo:2020qoa,Lobo:2021yem} through the search for signatures in particle accelerators and cosmic rays.
\subsubsection{Difficulties}

{\bf Absence of massless rainbow Finsler metric}. 
The Finsler approach had emerged as an opportunity to describe in a consistent way the intuition that the quantum spacetime probed by a high-energy particle would present some energy-momentum (of the particle itself) corrections, which is justified by different approaches to quantum 
gravity~\cite{Assanioussi:2014xmz,Weinfurtner:2008if}. Since then, proposals of rainbow metrics have considered a smooth transition from massive to  massless cases, not only from the point of view of the trajectories, but from the metric itself. This is not the case for the Finsler approach presented here. Although the trajectories and symmetries are defined for both massive and massless cases by considering the $m\rightarrow 0$ limit, the rainbow metric of Finsler geometry, given by Equation~\eqref{f-rainb}, 
is certainly 
not 
defined for massless particles. The reason for this is the point 
that when passing from the Hamiltonian to the Lagrangian formalism, we defined an arc-length functional, which is not a legitimate action functional for massless particles. 
In other words, 
a crucial step for deriving the Finsler function is the handling of the Lagrange multiplier $\lambda$ of action \eqref{action-ham1}, which can only be solved if the particle is massive, as can be 
found 
in~Refs. 
\cite{Amelino-Camelia:2014rga,Lobo:2016xzq,Lobo:2016lxm,Lobo:2020qoa}. A possibility that has been explored consisted of not solving the equation for $\lambda$ and defining a metric that depends on $\lambda$ and on velocities from a Polyakov-like action for free particles (instead of the Nambu--Goto one given by the 
arc-length), 
which turned out to be out of the Finsler geometry scope~\cite{Lobo:2016xzq,Lobo:2016lxm}. However, this possibility has not been further explored beyond preliminary investigations. The issue of the absence of a massless rainbow-Finsler metric could be circumvented by proposing a different kind of geometry, which from the very beginning started from the momenta formulation, like the other possibility described in this 
paper, namely the 
 Hamilton 
geometry.
\par
{\bf Definition only through perturbations}. 
 the  
Finsler geometry has been considered in this paper 
in this context at most perturbatively around the quantum-gravity-length scale (or inverse of energy 
scale), which may be considered as a negative point if one aims to 
 make 
it at a more fundamental or theoretical level. Nevertheless, from the pragmatic perspective of phenomenology, since such effects, if they exist, are minute, then the perturbative approach is enough for proposing new effects that could serve as avenues of experimental investigation.
\par
{\bf The handling of finite symmetries}. 
Another issue that can be problematic is the handling of finite symmetries in the 
Finsler context. 
 Up to 
today, 
 the  
connection between Finsler geometry and quantum gravity phenomenology has not faced the issue of integrating the symmetries and finding finite versions of deformed Lorentz transformations. Some initial investigations were carried out 
in~Ref. 
\cite{Lobo:2021yem} from the momentum space perspective, but further 
issues 
are being currently faced by some authors of the present paper.
\subsection{Hamilton Geometry}
The descriptions of the points in this section will be a bit shorter than the previous ones, because some universal points we already described above; therefore, we instruct the reader to check on them when that is the case.

\subsubsection{Advantages}

{\bf Presence of a massless rainbow Hamilton metric}. 
Differently from the Finsler case, the 
Hamilton geometry does not need an arc-length functional; 
instead, it only needs a given Hamiltonian, from which the metric, non-linear connection, and symmetries are derived. This means that from the very beginning, the massless limit of geometrical quantities exists.
\par
{\bf Does not require perturbative methods}. 
Another positive point about the Hamilton geometry is the finding 
that one can handle with 
the exact form of the proposed Hamiltonian, and it does not need to consider perturbations around a certain scale. Instead, independently of the form of the (smooth) dispersion relation that arises from {de facto} approaches to quantum gravity, the geometry can be handled, as has been 
considered, e.g.,
 in Refs. 
 \cite{Barcaroli:2016yrl,Barcaroli:2017gvg}.
\par
{\bf Preservation of the 
 relativity principle 
and the handling of symmetries}. 
Due to the proximity of this approach 
 to  
the way that the DSR formalism generally 
handles with Planck scale corrections, i.e., from the point of view of momentum space and Hamilton equations, the handling of symmetries is facilitated in this approach. For instance, it is straightforward to find the conserved charges from the killing vectors, which generate finite transformations that are momenta-dependent without tedious terms in the denominator of the equations when one is working in velocity space, as Finsler geometry is initially formulated (or without mass terms in the denominator in the Finsler version of the phase space).
\par
{\bf Generalization to curved spacetimes}. 
This approach is 
considered in more curved space 
cases, 
beyond the $q$-de Sitter exemplified  in this paper; for instance, its spherically symmetric and cosmological versions were 
explored 
giving rise to interesting phenomenological opportunities, from the point of view of time delays and gravitational redshift, among others (for some applications of Hamilton geometry in this context, 
see~\cite{Pfeifer:2018pty} and references therein).

\subsubsection{Difficulties}

{\bf Non-geodesic trajectory}. 
An issue that 
 may be considered  
problematic is the point 
that the worldlines of particles, given by the Hamilton equations, are not geodesics of the non-linear connection 
that  
means 
that there exists a force term in the geodesic equation, which is in contrast with the Finsler case. This is a property of 
the Hamilton geometry, as has been shown 
in~Ref. 
\cite{Barcaroli:2015xda}, and is not specific to the $q$-de Sitter case analyzed here.
\par
{\bf Absence of the arc-length}. 
The 
Hamilton geometry does not dwell with an arc-length functional that 
means that the only geodesics present 
are 
those of the non-linear or of the $N$-linear connections 
and there are no extremizing ones. The absence of a function that allows one to measure distances in spacetime can be seen as a difficulty of this 
geometry; 
if 
the 
distances cannot be calculated, 
one could wonder what  such a metric means. Even if 
the norm of a tangent vector can be integrated, 
this integral would not be, in general, parametrization-independent, which is also a drawback of this tentative. 
Besides, 
 the absence of 
an arc-length  
limits the phenomenology of the preservation of the clock postulate that was discussed in the Finsler case.


\section{Final Remarks}\label{sec:conc}

We revised two proposals that have been considered as candidates for describing the quantum configuration and phase spaces probed by particles whose kinematics are modified by a length scale identified as the quantum gravity scale. 
\par
Finsler geometry starts from a configuration space framework that presents applications on its own in biology, thermodynamics, and modified gravity; and it finds a natural environment in quantum gravity phenomenology due to its power to describe a scenario in which important principles that guided physics in the XXth century, such as the relativity principle, are preserved even at a Planckian regime. Besides its traditional description in terms of the couple spacetime and velocity 
space (configuration space), we also explored its development in terms of the induced couple spacetime 
and momentum 
space (phase space), which is actually more appropriate for a pure quantum description than the configuration space. Some points
that we consider positive and negative and 
 which 
are consequences of the requirements for using the Finsler language, 
 the derivation of an arc-length functional defined in the slit tangent bundle, 
are 
discussed in Section~\ref{sec:advdiff}. 
\par
The second case of the present 
study 
is the  
Hamilton geometry, whose properties are derived directly from the Hamiltonian itself, without the need to go through the definition of an 
arc-length.  
 Actually,  
in general, the Hamilton metric does not even define a curve-parametrization-invariant length measure 
which brings some limitations to phenomenological investigations of this subject in quantum gravity. On the other hand, 
this issue 
circumvents some intrinsic difficulties of Finsler geometry, which were also discussed in Section~\ref{sec:advdiff}.
\par
The 
goal of this paper was 
to review some topics of these two important 
geometries by using kinematical descriptions of particles whose behavior might present departures from special relativity results due to the effective quantum gravity influence. We also aimed to bring some points that we consider as under-explored perspectives on the subject by explicitly presenting some geometric quantities that are dual to 
those, 
in which 
 those quantities  
were originally presented, such as the dual metrics and non-linear connections (whose Finslerian one was proposed in this paper, by inspiration of definitions in the Hamilton geometry literature) of Finsler and Hamilton geometries in the cotangent and tangent bundles, respectively.
\par
{At least two global points could be considered insufficiently explored or unexplored in this subject. One is the geometry probed by an (non-)interacting multi-particle system. Some challenges of this problem can be found, for instance, 
in~Ref. 
\cite{Hossenfelder:2014ifa}, but the relations between the approaches there described and Finsler/Hamilton geometries remains unclear. Another point that remains unexplored consists of the dynamics of the configuration/phase space in a way that is compatible with quantum gravity phenomenology-inspired approaches. For instance, one could wonder if there exists a gravitational field theory defined in Finsler or Hamilton spaces that has $q$-de Sitter or other proposals as solutions, and how matter would interact in this scenario. 
The 
exploration of this topic might shed light on 
 the one regarding a multi-particle system. 
These are more 
challenges that might  
be 
subjects of the future research 
in this area 
 and which may help 
to build a bridge between quantum  and modified gravities. 
}





\vspace{6pt} 




\section*{Ackowledgements}I.P.L. was partially supported by the National Council for Scientific and Technological Development---CNPq grant 306414/2020-1, and by the grant 3197/2021, Para\'iba State Research Foundation (FAPESQ). 
I.P.L. would like to acknowledge the contribution of the COST Action CA18108. 
L.C.N.S. would like to thank Conselho Nacional de Desenvolvimento Cient\'ifico e Tecnol\'ogico (CNPq) for partial financial support through the research project no. 164762/2020-5. 
V.B.B. is partially supported by CNPq through the research project no. 307211/2020-7. 
P.H.M. and S.A. thank Coordena\c c\~ao de Aperfei\c coamento de Pessoal de N\'ivel Superior---Brazil (CAPES)---Finance Code 001, for financial support.  G.V.S thank Conselho Nacional de Desenvolvimento Cient\'ifico e Tecnol\'ogico (CNPq) for financial support. 
E.R. and G.M were supported by the PIBIC program of the Federal University of Para\'iba.


\appendix
\section[\appendixname~\thesection]{Dual Finsler Nonlinear Connection}\label{app:conn}
The momenta of a particle in Finsler geometry, given by the following expression, 
\begin{equation}
    p_{\mu}=m\frac{\partial F}{\partial y^{\mu}}\equiv 
m\dot{\partial}_{\mu}F\, ,
    \end{equation}
defines a kinematical map between velocity and momenta variables at each given point in the base manifold $M$. We refer to such a map as 
\begin{align}
    &\flat: \quad \widetilde{TM}\rightarrow \widetilde{T^{\ast}}M\\
    &(x,y)\mapsto \flat(x,y)=(x,m\dot{\partial}F(x,y))=(x,p(x,y))\, .
\end{align}

Inspired by the construction 
of~Ref. 
\cite{Barcaroli:2015xda}, the condition that a nonlinear connection in the tangent bundle is dual to one in the cotangent bundle by a 
kinematical map,  
$\flat$,  
is that such an application maps the tangent space of the tangent bundle onto the tangent space of the cotangent bundle. This means that the differential of such a map maps the preferred basis of one tangent 
space, 
$\delta_{\mu}=\partial_{\mu}-N^{\nu}{}_{\mu}\dot{\partial}_{\nu}$,  
onto the other,  
$d\, \flat (\delta_{\mu})=\delta'_{\mu}=\partial_{\mu}-O_{\mu\nu}\dot{\partial}^{\nu}$. This means that the action of this differential on a vector $X=X^{\mu}\partial_{\mu}+\dot{X}^{\mu}\dot{\partial}_{\mu}$ is given 
by 
\begin{align}
    d\, \flat_{(x,y)}:\quad T_{(x,y)}\widetilde{TM}&\rightarrow T_{\flat(x,y)}\widetilde{T^*M}\, ,\\
    X=X^{\mu}\partial_{\mu}+\dot{X}^{\mu}\dot{\partial}_{\mu}&\mapsto d\, \flat_{(x,y)}(X)=X^{\mu}d\, \flat_{(x,y)}(\partial_{\mu})+\dot{X}^{\mu}d\, \flat_{(x,y)}(\dot{\partial}_{\mu})\\
    &=X^{\mu}(\partial_{\mu}+m\partial_{\mu}\dot{\partial}_{\nu} F\bar{\partial}^{\nu})+m\dot{X}^{\mu}\dot{\partial}_{\mu}\dot{\partial}_{\nu}F\bar{\partial}^{\nu}\, .
\end{align}

By acting on the basis vectors $\delta_{\mu}=\partial_{\mu}-N^{\nu}{}_{\mu}\dot{\partial}_{\nu}$, 
one finds: 
\begin{align}
    d\, \flat_{(x,y)}(\delta_{\mu})=d\, \flat_{(x,y)}(\delta_{\mu})-N^{\nu}{}_{\mu}d\, \flat_{(x,y)}(\dot{\partial}_{\nu})=\partial_{\mu}+m\partial_{\mu}\dot{\partial}_{\nu} F\bar{\partial}^{\nu}-m N^{\nu}{}_{\mu}\dot{\partial}_{\nu}\dot{\partial}_{\alpha}F\bar{\partial}^{\alpha}\, .
\end{align}

In order to simplify this expression, 
the relation 
$2g_{\nu\alpha}=\dot{\partial}_{\nu}\dot{\partial}_{\alpha}F^2=\dot{\partial}_{\nu}(2F\dot{\partial}_{\alpha}F)$ 
is used that leads to 
\begin{equation}
    \dot{\partial}_{\nu}\dot{\partial}_{\alpha}F=\frac{g_{\nu\alpha}-p_{\nu}p_{\alpha}/m^2}{F}\, .
\end{equation}

From this expression, 
one finds 
that 
\begin{equation}
    d\, \flat_{(x,y)}(\delta_{\mu})=\partial_{\mu}-m\left[N^{\alpha}{}_{\mu}\frac{(g_{\alpha\nu}-p_{\alpha}p_{\nu}/m^2)}{F}-\partial_{\mu}\dot{\partial}_{\nu}F\right]\bar{\partial}^{\nu}=\partial_{\mu}+O_{\mu\nu}\bar{\partial}^{\nu}\, ,
\end{equation}
which leads to the dual nonlinear connection, 
\begin{equation}
    O_{\mu\nu}(x,p)=-m\left[N^{\alpha}{}_{\mu}\frac{(g_{\alpha\nu}-p_{\alpha}p_{\nu}/m^2)}{F}-\partial_{\mu}\dot{\partial}_{\nu}F\right]\Bigg{|}_{(x,y(p))}\, .
\end{equation}






\bibliographystyle{utphys}
\bibliography{review-fh}

\end{document}